\documentclass[apj,onecolumn]{emulateapj}
\input epsf.tex
\begin{document}
\title{On the Absorption and Redistribution of Energy in Irradiated Planets} 
\author{Brad M. S. Hansen\altaffilmark{1}
}
\altaffiltext{1}{Department of Physics \& Astronomy and Institute of Geophysics \& Planetary Physics, University of California Los Angeles, Los Angeles, CA 90095, hansen@astro.ucla.edu}


\lefthead{Hansen }
\righthead{Irradiated Toy}

\begin{abstract}
We present a sequence of toy models for irradiated planet atmospheres, in which the effects of geometry and energy
redistribution are modelled self-consistently. We use separate but coupled grey atmosphere models to treat the
ingoing stellar irradiation and outgoing planetary reradiation. We investigate how observed quantities such
as full phase secondary eclipses and orbital phase curves depend on various important parameters, such as the
depth at which irradiation is absorbed and the depth at which energy is redistributed. We also compare our
results to the more detailed radiative transfer models in the literature, in order to understand how those
map onto the toy model parameter space. Such an approach can prove complementary to more detailed calculations,
in that they demonstrate, in a simple way, how the solutions change depending on where, and how, energy redistribution occurs. As an example of the value of such models, we demonstrate how energy redistribution and temperature equilibration at moderate optical
depths can lead to temperature inversions in the planetary atmosphere, which may be of some relevance to recent
observational findings.
\end{abstract}


\section{Introduction}

The past several years have seen a rapid and exciting increase in the amount of
information available concerning the physical properties of extrasolar planets. In
particular, the class of short orbital period planets, known informally as `Hot Jupiters',
has begun to yield information through a variety of observational channels. The discovery
of systems in which the orbital plane is almost edge-on to the line of sight has resulted
in the detection of planetary transits (Charbonneau et al. 2000; Henry et al. 2000;
 Bouchy et al. 2005). Optical and ultraviolet photometry and spectroscopy of such systems
can yield information about the absorbative properties of the planet atmosphere (Brown et al. 2001;
Charbonneau et al. 2002; Vidal-Madjar et al. 2003; Vidal-Madjar et al. 2004). In addition, the
unparalleled capabilities of the Spitzer Space Telescope has allowed the detection of
thermal infrared emission (Charbonneau et al. 2005; Deming et al. 2005,
Deming et al. 2006; Grillmair et al. 2007; Richardson et al. 2007) during secondary eclipses as
well as the measurement of phase curves for non-transitting systems (Harrington et al. 2006).

This flurry of observational activity has spurred a similar level of activity on the theoretical
front, with several independent models being used to infer the physical properties of the
planets from the observations (Seager \& Sasselov 2000; Barman, Hauschildt \& Allard 2001;
Cho et al. 2003; Menou et al 2003;
Sudarsky, Burrows \& Hubeny 2003; Burrows, Hubeny \& Sudarsky 2005; Dyudina et al. 2005; 
Cooper \& Showman 2005;
Seager et al. 2005; Barman, Hauschildt \& Allard 2005; Fortney et al. 2005; Fortney 2005; 
Burrows, Sudarsky \& Hubeny 2006; Cooper \& Showman 2006; Fortney et al. 2006; Langton \& Laughlin 2007). 
The resulting
inferences are not always entirely in agreement. As an example, consider each group's response
to the first detection of secondary eclipses. Burrows et al. (2005) concluded that the
redistribution of energy from the substellar to the antistellar face of the planet was
weak. On the other hand, Barman, Hauschildt \& Allard (2005), as well as Fortney et al. (2005),
concluded that the redistribution was strong, while the opinion of Seager et al. (2005) fell
somewhere in between. Similarly, predictions for the effects of atmospheric fluid motions 
on the emergent intensity pattern differ substantially from group to group (Cho et al. 2003; Cooper \&
Showman 2006; Langton \& Laughlin 2007).

This diversity of opinion is hardly surprising, given the complexity of the problem. The study
of Hot Jupiter atmospheres requires that we understand the atmospheric chemistry, radiative transfer
and hydrodynamics of the planetary atmosphere. These are affected by the thermal structure and
evolution of the planet, by the effects of photochemistry induced by ultraviolet light from the star,
the possible existence and behaviour of atmospheric condensates, whether they form clouds, and
whether (or if) energy is redistributed across the surface of the planet or reradiated in situ.
Furthermore, there are a variety of different observational probes, each of which provides information
but in ways that differ, sometimes subtly, from other methods. The complexity of the coupled radiative
transfer, chemical equilibrium and atmospheric flow models also restrict somewhat the flexibility of
the models in providing the true qualitative or geometric information necessary to develop a real
understanding of what the observations are telling us.

The overall intent in the pages to follow is pedagogical. Clearly, detailed model atmospheres are necessary to
fully interpret the emerging observations. However, simplified models that focus on key physical elements
 can be useful to try and
develop a qualitative understanding of what elements of the model each of the
 various extant observations is actually probing. To do this, we construct a sequence of 
analytic and semi-analytic toy models. These models are chosen to be simple enough to be transparent
while still capturing the qualitative features of the more detailed models. The hope is that one can
use this approach to tailor our interpretations of the more detailed calculations to reach a 
proper synthesis between theory and the different observations that we have at our disposal.
In \S~\ref{Nada} we review our simplest
model, based on a variation of the grey atmosphere model. In this first case we will
assume no redistribution of energy across the face of the planet. In \S~\ref{Apply} we
then examine how the various available observations relate to the underlying structure
of the model. In \S~\ref{Finally} we then introduce the physics of energy redistribution
into the model and examine how this affects the observations in \S~\ref{MoreApps}. Finally,
in \S~\ref{Finis}, we place our models within the context of both observations and more
detailed models in the literature, and examine possible implications for future generations
of models.
 
\section{Models without Redistribution}
\label{Nada}

To begin, we briefly review the basic steps in calculating the grey atmosphere model. These
may be found in almost any textbook on radiative transfer (e.g. Mihalas 1978) but it is worth
reviewing them again as we will need to employ some subtle variations in subsequent sections. 
In doing so we will also derive a simple model which gives an explicit analytic description 
of the current philosophy in the literature for treating this situation. This will serve as
a useful baseline for comparison in later sections.

\subsection{The Simplest Model}
\label{Model0}

In the grey approximation, we work with a frequency-integrated version of the radiative transfer 
equation
\begin{equation}
\mu \frac{\partial I}{\partial \tau} = I - S \label{RadTrans}
\end{equation}
where $I$ is the specific intensity, and $S$ is the source function. $I$ is a function of optical
depth $\tau$ and angle cosine (relative to the normal direction) $\mu$. In local thermodynamic equilibrium,
where the optical depth is due to absorption rather than scattering, $S = B$, the frequency-integrated
Planck function. Integration over angle yields the equation for flux conservation, which is satisfied as
long as the conditions for radiative equilibrium are met, i.e. $J=B$. $J$ is the mean intensity, the
zeroth moment of $I$ over angle. The first moment of equation~(\ref{RadTrans}) yields 
\begin{equation}
\frac{\partial K}{\partial \tau} = H \label{FirMom}
\end{equation}
where $H$ and $K$ are the first and second moments of $I$. At this point, we adopt the Eddington approximation, assuming that 
$J=3 K$. This results in the equation
\begin{equation}
3 \frac{\partial^2 J}{\partial \tau^2} = J - B = 0   \label{J0}
\end{equation}
which yields a solution for $J(\tau)$, $J = J_0 + J_1 \tau$. 

The constant $J_1$ is determined from equation~(\ref{FirMom}) and the lower boundary condition on the flux
 $F_{int} = 
4 H$ entering
the atmosphere from below. In the terminology of Mihalas (1978), $F$ is the ``Astrophysical Flux'' and
$H$ is the ``Eddington Flux''.
 Since $S = B = J$, one can also determine $I$ from the formal solution to the first
order differential equation~(\ref{RadTrans})
\begin{equation}
 I(0,\mu) = \int_0^{\infty} \left[ J_0 + J_1 t \right] e^{-t/\mu} \frac{dt}{\mu} = J_0 + \frac{3}{4} F_{int} \mu.
\end{equation}
The final act is to determine the constant $J_0$ by the calculation of the emergent flux at the surface
($\tau=0$) from the
expression for $I$ and the condition that it must equal $F_{int}$
\begin{equation}
F_{int} = 2 \int_{-1}^{1} \mu I(0,\mu) d\mu = 2 \int_{0}^{1} \mu I(0,\mu) d\mu + 2 \int_{-1}^{0} \mu I(0,\mu) d\mu.
\end{equation}
For most astronomical applications, the
 second term here is  set to zero as it represents the contribution from rays directed inward at the
surface, and most stars are not subject to any significant level of irradiation.
 However, in the case of Hot Jupiters, this particular term is of great importance as it introduces the irradiation from
the parent star into
the problem. The simple approximation used most often in the literature is uniform irradiation i.e.
$I(0,\mu) = I_0$ for $\mu<0$, although the reality of the situation is that all the incoming rays are
coming from a single direction, that of the star. Nevertheless, if we proceed with the uniform irradiation
approximation, we can use this
to determine $J_0$ and derive  the final solution for the specific intensity
\begin{equation}
I(0,\mu) = \frac{3}{4} F_{int} \left( \mu + \frac{2}{3} \right) + f  I_0,  \label{Solution1}
\end{equation}
where we have included a dilution factor $f$ in the spirit of the more sophisticated modellers.
This is how the detailed models attempt to account for the difference between the true situation and
the isotropic irradiation assumption.
 Because
of the computation expense involved, these models are still calculated in the one dimensional case, but
the incoming radiation is `diluted' by $f$ in order to approximately mimic the effect of averaging over
the true non-uniform irradiation pattern.

An important point to note is that the {\em net flux} through any given point in the model atmosphere is still
only the internally generated $F_{int}$, since no energy from the irradiation is permanently absorbed by
the planet in this baseline model -- it is all reradiated. 
However, the flux the observer sees  
 is the sum total of all flux travelling outwards from the planetary surface, and so includes the contribution
of reradiated stellar light. In the case of a secondary eclipse, this is radiated from the 
surface in the direction of the star. (For this simple model the distinction is moot, but it will become
important in subsequent sections).
\begin{equation}
F_{full} = 2 \int_0^1 \mu I(0,\mu) d\mu = F_{int} +   f I_0. \label{SupCon1}
\end{equation}

The traditional assumption found in the literature has been to adopt a value
of $f=1/2$ to represent the case of no energy redistribution
over the surface of the planet.
 In this case
the dilution is meant to represent the fact that the incoming flux is absorbed
from plane parallel rays spread over  over a surface $\pi R_p^2$ (where $R_p$ is the planet radius) but
is reradiated isotropically from each point on the irradiated hemisphere and
so the energy is emitted into a solid angle of $2 \pi$.
If the energy is redistributed uniformly over
the entire planet by fluid motions/winds and only then reradiated, then the angle is $4 \pi$ and
the dilution factor is $f=1/4$.

\subsection{A Better No-Redistribution Model}
\label{Model1}

The previous section gave an explicit example for how irradiated planets are currently treated using
a one-dimensional model and a dilution factor. However, we can adapt this formalism easily to treat
the geometry of the no-redistribution model properly. 

The first improvement to make is to dispense with the assumption of an isotropic irradiation. Consider
instead the irradiation is due to a mono-directional beam\footnote{The readers who are concerned about
the use of the Eddington approximation in concert with a monodirectional beam should defer their
skepticism until \S~\ref{Model2}. The discussion here is illustrative and this point will be addressed
further when we come to the actual model we will use.},
 so that $I(0,\mu) = I_0 \delta(\theta + \theta_0) \delta(\phi-\phi_0)$
for $\mu<0$, and $\theta$ and $\phi$ are the poloidal and toroidal co-ordinates for a given
point on the planetary surface. Furthermore, we can now choose $\theta_0$ \& $\phi_0$ so that, for each point on the planet's substellar
hemisphere, the irradiating beam is coming from the direction of the star. To achieve this, we simply note
that $\mu_0$ is the angle cosine between the present location and the substellar point.
To determine the constant $J_0$ in this case, we need to match to the outgoing flux $F_{int} + \mu_0 I_0$. 
With this change, equation~(\ref{Solution1}) is replaced by
\begin{equation}
I(0,\mu_0,\mu) = \frac{3}{4} F_{int} \left( \mu + \frac{2}{3} \right) + \mu_0 I_0 \label{Solution2}
\end{equation}
We omit $f$ in this expression because it is no longer
needed. 
This is now a solution that varies across the surface of the irradiated planet, as determined by the
local angle of incidence of the irradiation. 
To obtain the flux we actually observe, we simply integrate over $\theta$ and $\phi$.
Evaluating the integral at full phase (secondary eclipse), we find 
\begin{equation}
F_{full} = 2 \int_0^1 \mu_0 I(0,\mu_0,\mu_0) d\mu_0 = F_{int} + \frac{2}{3} I_0.
\end{equation}
It is of interest to note that, in order to make equation~(\ref{SupCon1}) match this expression, we
require $f = \frac{2}{3}$, rather than the value $f=\frac{1}{2}$ that is normally quoted. This is
because our new model does not have a uniform distribution of flux over the surface of the planet -- rather, it is hotter near the substellar point because that is where the irradiation is the most intense. Thus, the uniform dilution approximation underestimates the secondary eclipse flux. 

We can easily confirm global energy conservation in this model. With the monodirectional beam of irradiation,
the total flux impinging on the surface of the planet is 
\begin{equation}
F_{irr} = 2 \int_0^1 d\mu_0 \left( \int_{-1}^0 \mu I_0 \delta(\mu + \mu_0) d\mu\right) =  I_0.
\end{equation}
Using equation~(\ref{Solution2}) to calculate the total emitted flux returns the same value.
In the case of the isotropic irradiation model, 
the total flux emitted from the planet
may be expressed as
\begin{equation}
2 \int_0^1 d\mu_0 \left( 2 \int_0^1 \mu I(0,\mu) d\mu \right) = 2 F_{int} + 2 f I_0
\end{equation}
if we use equation~(\ref{Solution1}). The equality of this with $F_{irr}$ is how we arrive at the
value $f=1/2$. 

Thus, our very first conclusion, even from a relatively simple model, is that the detailed models in
the literature often dilute even their `no-redistribution' models too much. Models for secondary
eclipses should not be too democratically averaged over the surface of the planet, but should be weighted
in favour of the hotter substellar point. This is in agreement with a similar comparison performed by
Barman, Hauschildt \& Allard (2005), who compared two dimensional no redistribution models with the
same one dimensional models for $f=1/2$ (their ``$\alpha$'' is equivalent to our $f$). 

\subsection{A Two Opacity Model}
\label{Model2}

We wish to improve our toy model still further, because there is an important piece of physics
still missing.
The wavelengths of the incoming stellar 
photons are very different from those of the reradiated photons. This is a consequence of the 
very different temperatures and hence spectral energy distributions of the star and the planet.
The result is that the opacities and hence optical depths can be very different for the energy
flow in the optical and infrared wavelength regimes. Thus, we will modify our model such that we 
treat the absorption of stellar irradiation seperately and with a different optical depth than
we do the subsequent reradiation.

In order to do this, let us treat the grey version of the radiative transfer equation in a `two-stream'
fashion, with two specific intensities, $I_1$ to describe the optical radiation, and $I_2$ to describe
the infrared radiation. This allows us to treat the two streams with different opacities and emissivities.
In this limit, we have two equations of radiative transfer, corresponding to the optical and infra-red
respectively.
\begin{eqnarray}
\mu \frac{\partial I_1}{\partial \tau_1} & = & - I_1 \\
\mu \frac{\partial I_2}{\partial \tau_2} & = & S_2 - I_2
\end{eqnarray}
where we have set $S_1$, the source function in the optical, to zero, since we assume the planet
has negligible emissivity in the optical. Furthermore, the constraint of radiative
equilibrium requires that all the energy removed from the beam (in both wavelength regimes)
is replaced by the infrared source function $S_2$, so that
\begin{equation}
\kappa_1 J_1 + \kappa_2 J_2 = \kappa_2 S_2,
\end{equation}
where $\kappa_i$ and $J_i$ are the appropriate opacities and mean intensities\footnote{This approach bears a 
superficial resemblance to the formalism of Hubeny, Burrows \& Sudarsky (2003), in which the two opacities are
derived from different frequency averages. However, that study depends critically on the frequency and temperature
dependance of the opacity microphysics, as it was focussed on the possibility that multiple solutions might 
arise for a given physical situation. 
The multiple solutions derived by Hubeny et al. do
not arise in our case, because we assume the opacity ratio $\gamma$ is a constant.}. Thus
our radiative equilibrium expression is slightly modified from the usual form,
\begin{equation}
S_2 = J_2 + \frac{\kappa_1}{\kappa_2} J_1 = J_2 + \gamma J_1. \label{RadEq}
\end{equation}

Within this approximation, the first moment equations become
\begin{eqnarray}
\frac{\partial H_1 }{\partial \tau_1} & = & - J_1 \\
\frac{\partial H_2 }{\partial \tau_2} & = & S_2 - J_2.
\end{eqnarray}
Using equation~(\ref{RadEq}), and $\tau_1 = \gamma \tau_2$, we can combine these
equations into
\begin{equation}
\frac{\partial H_2}{\partial \tau_2} = -\frac{\partial H_1}{\partial \tau_2}. \label{Hbeam}
\end{equation}
This amounts to the rather obvious statement that all the flux removed from the incoming
beam in the optical must emerge in the infrared beam.

The attenuation of the incident beam is described by  
\begin{equation}
 I(\tau,\mu)=I(0,\mu) e^{\gamma \tau/\mu},
\end{equation}
where we have replaced $\tau_2$ with $\tau$, our default optical depth scale henceforth.
Recall that $\mu<0$, since this is an inwardly directed beam.
The local flux (travelling inwards) is
\begin{equation}
F = \int_0^{2\pi} \delta (\phi - \phi_0) \int_{-1}^{0} \mu I_0 \delta (\mu + \mu_0) e^{\gamma \tau/\mu} d\mu = -  \mu_0 I_0 e^{-\gamma \tau/\mu_0}.\label{Fbeam}
\end{equation}
This is the flux flowing inwards due to the irradiation. We have thus a measure of how much energy is deposited
at each depth. We assume the energy is absorbed, thermalised and now reradiated in the infrared.
This is done by incorporating equation~(\ref{Fbeam}) into equation~(\ref{Hbeam}), 
\begin{equation}
\frac{\partial H}{\partial \tau} =  - \frac{\gamma}{4} I_0 e^{-\gamma \tau/\mu_0}.
\end{equation}
The gradient in $H$ is negative, because the total energy flux carried outwards by the
radiation field is less as one gets deeper into the planet.

We once again make the Eddington approximation. This 
is a poor one for a monodirectional beam, but we note that the approximation is only
applied here to the reradiation field. 
Once the energy is thermalised and
included in the reradiation energy flow,
 which is much more isotropically distributed, this is a more reasonable
 approximation.
Thus we obtain the integrable equation
\begin{equation}
\frac{\partial^2 J}{\partial \tau^2} = - \frac{3}{4} \gamma I_0 e^{-\gamma \tau/\mu_0}
\label{J1}
\end{equation}
which can be integrated to yield
\begin{equation}
 J = J_0 + J_1 \tau - \frac{3}{4} \frac{\mu_0^2 I_0}{\gamma} e^{-\gamma \tau/\mu_0}.
\end{equation}

The integration constant $J_1$ is determined, as before, by the internal flux emerging from the
planet ($ J_1 = \frac{3}{4} F_{int}$). To determine the other constant, we need to
ensure the total flux emerging at the top of the atmosphere reflects the appropriate
sum of internal and absorbed flux. To that end, we calculate the specific intensity
as a function of angle at the surface
\begin{equation}
I(0,\mu_0,\mu) = \int_0^{\infty} J e^{-t/\mu} \frac{dt}{\mu} = J_0 + \frac{3}{4} F_{int} \mu
- \frac{3}{4} \frac{\mu_0^2 I_0}{\gamma (1 + \gamma \mu/\mu_0)}.
\end{equation}

One may then integrate this over angle to determine the total flux emerging from the surface
\begin{equation}
F = 2 \int_0^1 \mu I(0,\mu) d\mu = J_0 + \frac{1}{2} F_{int} - \frac{3}{2} \frac{\mu_0^3 I_0}{\gamma^2}
+ \frac{3}{2} \frac{\mu_0^4 I_0}{\gamma^3} \ln ( 1 + \frac{\gamma}{\mu_0} ).
\end{equation}

This should match the sum of the internal radiation and flux incident on the top of the
atmosphere from the star, $F_{int} + \mu_0 I_0$, so that the final expression for the
emergent specific intensity is
\begin{equation}
I(0,\mu_0,\mu) = \frac{3}{4} F_{int} \left( \mu + \frac{2}{3} \right)
+  \mu_0 I_0 \left[ 1 - \frac{3}{4} \frac{\mu_0/\gamma}{1 + \gamma \mu/\mu_0}
+ \frac{3}{2} \left( \frac{\mu_0}{\gamma} \right)^2
- \frac{3}{2} \left( \frac{\mu_0}{\gamma} \right)^3 \ln \left( 1 + \frac{\gamma}{\mu_0} \right)
\right]. \label{Final_Solution}
\end{equation}
This expression now will allow us to calculate the emergent radiation field in a manner
appropriate to any particular observation geometry, given a particular choice of observer
orientation and orbital phase. The angle $\mu_0$ measures the angle for each point on the
surface relative to the illuminating star, while $\mu$ measures the angle relative to the
observer. The parameter $F_{int}$ is the internal heat flux emerging from the planet itself,
while $I_0$ encodes the strength of the irradiation. The ratio of $I_0$ relative to $F_{int}$ 
will determine whether the observational appearance is dominated by the reradiation of absorbed
energy.
The one other factor in this equation is $\gamma$. This is the ratio of the optical depths
for absorbed and reradiated energy.
It can be tuned to reflect an optical depth for incoming stellar light that
is either more ($\gamma>1$) or less ($\gamma<1$) opaque than that for the emerging thermal
radiation from the planet. Note that this does not reduce to equation~(\ref{Solution2}) in the case
of $\gamma=1$, because the quantity $I$ now refers only to infra-red light, and so the flux
carried in the beam is not constant with depth (as it is in \S~\ref{Model1}).

 In the deep absorption limit ($\gamma \rightarrow 0$),
\begin{equation}
I(0,\mu_0,\mu) \rightarrow  \frac{3}{4} F_{int} \left( \mu + \frac{2}{3} \right)
+  \mu_0 I_0 \left[ \frac{1}{2} + \frac{3}{4} \mu + \frac{3}{8} \frac{\gamma}{\mu_0}
\left( 1 - 2 \mu^2 \right) \right] + O(\gamma^2)  
\end{equation}
which can be cast to zeroth order in terms of just an extra contribution to the internal flux, as one would
expect
\begin{equation}
I(0,\mu_0,\mu) \rightarrow \frac{3}{4} \left( F_{int} +  \mu_0 I_0 \right)
\left( \mu + \frac{2}{3} \right) + \frac{3}{8} \gamma I_0 \left( 1 - 2 \mu^2 \right) + O(\gamma^2).
\end{equation}

In the opposite ($\gamma \rightarrow \infty$) limit (absorption high in the atmosphere), we have
\begin{equation}
I(0,\mu_0,\mu) \rightarrow  \frac{3}{4} F_{int} \left( \mu + \frac{2}{3} \right)
+  \mu_0 I_0 \left[ 1 + \frac{3}{2} \left( \frac{\mu_0}{\gamma} \right)^2 \left( 1 - \frac{1}{2 \mu} \right) - \frac{3}{2} \left( \frac{\mu_0}{\gamma} \right)^3 \ln \frac{\gamma}{\mu_0} + O(1/\gamma^3) \right]
\end{equation}
The last term, despite the logarithmic function of $\gamma$, still falls off faster that $1/\gamma^2$,
so retaining only the leading term in $\gamma$ leaves
\begin{equation}
I(0,\mu_0,\mu) \rightarrow  \frac{3}{4} F_{int} \left( \mu + \frac{2}{3} \right)
+  \mu_0 I_0 \left[ 1 + \frac{3}{2} \left( \frac{\mu_0}{\gamma} \right)^2 \left( 1 - \frac{1}{2 \mu} \right)\right] + O(\ln \gamma/\gamma^3).
\end{equation}
So, for high altitude absorption, the zeroth order limit is the simple no-redistribution model
of equation~\ref{Solution2}.\footnote{Note that the apparent divergence of the second order term as
$\mu \rightarrow 0$ is misleading. One cannot maintain the limit $\gamma \rightarrow \infty$ at
the same time as $\mu \rightarrow 0$. One can verify this by taking the appropriate edge-on
limit of the full expression, which remains finite.}

Thus, equation~(\ref{Final_Solution}) presents us with a simple, yet flexible model that we
can use to try and understand how different potential observations probe the properties
of Hot Jupiter atmospheres.

\section{Some Applications}
\label{Apply}

One of the primary goals of this study is to provide a flexible framework in which to
understand how the observations probe different elements of the physics. So, let
 us now consider now how this model is reflected in the various kind of observations that
are becoming available.

\subsection{Secondary Eclipses}
\label{Sec1}

A secondary eclipse occurs when a planet in an edge-on orbit (as measured by the observer) 
passes behind the star. The size of the flux
decrement observed yields the flux emerging from the substellar side of the planet. More
specifically, it represents the flux observed at full phase, when $\mu = \mu_0$. Evaluating
the emergent flux integral at this phase, we get
\begin{equation}
F_{full} = 2 \int_0^1 \mu_0 I(0,\mu_0,\mu_0) d\mu_0 = F_{int} +
\frac{1}{2} I_0 \left[ 1 + \frac{1}{4 \gamma} + \frac{1}{\gamma^2} - \frac{3}{4 \gamma (1 + \gamma)}
+ \frac{\gamma}{2} - \gamma^2 - \frac{1}{\gamma^3} \ln (1 + \gamma) - \gamma^3 \ln (1 + \frac{1}{\gamma}) \right]. \label{OppF}
\end{equation}

In principle, comparing the observed $F_{full}$ to model values can tell us whether energy is being
redistributed to the antistellar side, because then the observed value would be less than expected from
the no-redistribution model. We can express this as an effective value of $f$ by dividing the reradiation
term by $I_0$, so that
\begin{equation}
f_{eff} =
\frac{1}{2}\left[ 1 + \frac{1}{4 \gamma} + \frac{1}{\gamma^2} - \frac{3}{4 \gamma (1 + \gamma)}
+ \frac{\gamma}{2} - \gamma^2 - \frac{1}{\gamma^3} \ln (1 + \gamma) - \gamma^3 \ln (1 + \frac{1}{\gamma}) \right]. 
\end{equation}

Despite the algebraic length of the expression, there is actually remarkably little variation 
as a function of $\gamma$. As $\gamma \rightarrow 0$, the limit is
\begin{equation}
f_{eff} \rightarrow \left[ \frac{17}{24} + \frac{3}{8} \gamma + O (\gamma^2) \right]
\end{equation}
while the limit as $\gamma \rightarrow \infty$ is
\begin{equation}
f_{eff} \rightarrow \left[ \frac{2}{3} + \frac{3}{5 \gamma^2} + O (\ln \gamma/\gamma^3)\right] .
\end{equation}
Figure~\ref{Opp} shows the function $f_{eff}(\gamma)$.
There is only a change of about 6\% between high and low values of $\gamma$.
We may thus conclude that, in case of no energy redistribution, it matters little where 
in the atmosphere the energy is actually absorbed (at least for secondary eclipse
measurements). 
We find that 
$f = 2/3$ in the large $\gamma$ limit, and in the low $\gamma$ limit,
$f = 17/24=0.708$. We note that this latter value corresponds to the same full-phase fraction
derived by Milne (1926) in his treatment of the `reflection effect' in close binaries. His
treatment amounts to a special case ($\gamma=1$) of the above treatment. The insensitivity of $f$ to
$\gamma$ is encouraging, 
because it suggests that the comparison of models to secondary eclipse fluxes is robust, in the
sense that any flux deficit is likely to indeed be due to the process of redistribution, and not
strongly affected by whether the energy is absorbed high in the atmosphere or deep.

\subsection{Limb Darkening}

Of course, there is much more information potentially available than just the flux at a single
orientation. To start, let us consider the planetary limb-darkening law. Neglecting the internal
flux term, at full phase we have
\begin{equation}
\frac{I(0,\mu_0,\mu_0)}{I(0,1,1)} = \mu_0 \frac{ 1 - \frac{3}{4} \frac{\mu_0}{\gamma (1+\gamma)}
+ \frac{3}{2} \left(\frac{\mu_0}{\gamma}\right)^2 - \frac{3}{2} \left( \frac{\mu_0}{\gamma} \right)^3
\ln ( 1 + \frac{\gamma}{\mu_0} )}{1 - \frac{3}{4} \frac{1}{\gamma (1+\gamma)}
+ \frac{3}{2} \frac{1}{\gamma^2} - \frac{3}{2} \frac{1}{\gamma^3} \ln ( 1 + \gamma )}
\end{equation}
which is shown in Figure~\ref{Limb} for three values of $\gamma=0.1$,1 $\&$10. Also shown is the normal
limb-darkening law for internally generated heat
\begin{equation}
\frac{I(0,\mu_0)}{I(0,1)} = \frac{3}{5} \left( \mu_0 + \frac{2}{3} \right),
\end{equation}
and the limb-darkening law derived by Milne (1926) in the two-beam approximation
\begin{equation}
\frac{I(0,\mu_0,\mu_0)}{I(0,1,1)} = \left( \frac{1 + 2 \mu_0}{3} \right)^2.
\end{equation}

We see that all the irradiation-driven profiles are much more limb-darkened than the
traditional internally generated profile. This is to be expected, since, near the limb,
the irradiation is being absorbed at oblique angles and so spread over more area. 
Absorption high in the atmosphere makes the limb even darker, but the effect is small
relative to the dominant effect of the irradiation geometry. The steep dependance on $\mu_0$
is another way to illustrate why the true value of $f$ is $f=2/3$ rather than $f=1/2$; the region near
the substellar point, which is also the hottest, contributes disproportionately to the
reradiated flux observed at full phase.

\subsection{Secondary Transit Shapes}

During secondary transit, the face of the planet is steadily eclipsed by the star. Thus, the
manner in which the total light from the system decreases and increases can, in principle,
tell us how light is distributed on the face of the planet. Figure~\ref{Trans} shows the shape
of the secondary
transit ingress (i.e. the drop in the system flux as the star occults the planet) for three
cases. In each case, we assume the occulting body is a sharp, vertical boundary.
 The solid line indicates a planet with the no-redistribution illumination pattern oriented
exactly edge-on. The dotted line is for the same model but now with an inclination of 2~degrees.
 The dashed line indicates a uniform hemisphere planet oriented exactly edge-on. Each case assumed
the planet radius was 0.1 times that of the star and that the orbital semi-major axis is 12 stellar
radii. Separating out inclination effects is relatively easy based on the duration of the eclipse,
but the measuring differences in illumination pattern by determining the shape of the ingress is
going to be quite difficult. The differences are only a few percent at most.
This is consistent with other studies concerned with using the shape of secondary transit to
infer the presence of non-uniform brightness (Williams et al. 2006; Rauscher et al. 2007a). 
Greater potential for constraining models is likely to be found in detecting
variations in the timing (Williams et al. 2006) or amplitude (Rauscher et al. 2007b) of maximum depth.

Overall then, it appears as though the properties of secondary eclipses are not strongly 
affected by the value of $\gamma$ in this model. This is encouraging, since that
means we can make robust inferences about the nature of energy redistribution.

\subsection{Phase Curves}

Another way to probe the asymmetry of the flux distribution emerging from the planet is
to measure the phase curve -- the variation in the observed brightness of the planet as
it orbits the star.
For the simplest case, let us consider the phase curve for a planet which has two
hemispheres, each of uniform brightness but at different levels. This amounts to a model
in the spirit of the default model in \S~\ref{Model0}. Furthermore, let us set the antistellar
side to zero brightness, since it is the difference in flux between sub- and antistellar sides
that is important.

A location on the surface of the sphere is identified by a unit vector defined in terms
of two angles, $\theta$ and $\phi$,
\begin{equation}
{\bf \hat{n}} = \left( \cos \phi \sin \theta, \sin \phi \sin \theta, \cos \theta \right).
\end{equation}
We consider the direction towards the observer to lie along the x axis, so that the
angle cosine between the observer and local normal on the surface of the sphere is then 
$\mu = \cos \phi \sin \theta$. We also need to specify the direction from which the irradiation
comes, using the unit vector
\begin{equation}
{\bf \hat{n}_*} = \left( \cos \lambda \cos i, \sin \lambda, -\cos \lambda \sin i \right).
\end{equation}
where the angle $\lambda$ can circulate in a plane rotated by an angle $i$ about the y-axis.
Thus, the angle cosine between this direction and local normal is
\begin{equation}
\mu_0 = {\bf \hat{n}.\hat{n}_*} = \cos \phi \cos \lambda \sin \theta \cos i + \sin \phi \sin \lambda \cos \theta - \cos \theta \cos \lambda \sin i
\end{equation}
To take account of the phase, we note that we need not account for the true orbital motion, but only
the rotation of the irradiation pattern, since the actual orbital displacement is unresolved by any applicable photometric
measurement. Thus, our phase curve is described by the circulation of $\lambda$. For the edge-on case
($i=0$) this has an analytic solution for the phase curve
\begin{equation}
F=\int_{-\pi/2}^{\lambda + \pi/2} d\phi \int_0^{\pi} \sin^2 \theta \cos\phi d\theta \label{PhaseInt}
\end{equation}
for $\lambda<0$. This applies for brightness of unity for $\mu_0 = \cos(\phi-\lambda)\sin \theta >0$
and zero otherwise. Evaluation of the integral yields
\begin{equation}
F = \frac{\pi}{2} \left[ 1 + \cos \lambda \right] \label{FlatSol}
\end{equation}
which also holds for $\lambda>0$. Figure~\ref{Phase12} shows the phase curves for this model and a range
of inclinations.

We can repeat the same exercise for the 
 model in \S~\ref{Model1},  except that the surface brightness is no longer uniform, since 
there is an extra factor of $\mu_0$ in the integrand in equation~(\ref{PhaseInt}). The solution is
then
\begin{eqnarray}
F & = & \frac{2 \pi}{3} \cos \lambda + \frac{2}{3} \left[ \lambda \cos \lambda - \sin \lambda \right] \, \, \, ( \lambda < 0 ) \nonumber \\
 & = & \frac{2 \pi}{3} \cos \lambda - \frac{2}{3} \left[ \lambda \cos \lambda - \sin \lambda \right] \, \, \, ( \lambda > 0 ).  \label{SpotSol}
\end{eqnarray}

The lower panel of Figure~\ref{Phase12} shows the appropriate phase curves. The qualitative 
difference is that this model has broader troughs and narrower peaks than the uniform hemisphere model. This is a consequence of the concentration of the illumination near the substellar point, so that the
phase curve is more sensitive to how this particular region is projected onto the line of sight. Some
readers will no doubt note that this is the same as the phase curve for a Lambert sphere (e.g. Russell 1916).
Although the Lambert sphere is usually invoked to describe reflected light, the conditions under which
it is derived (single, isotropic scattering) amount to the same conditions as applicable to thermal
reradiation.

In Figure~\ref{Phase3} we show a comparison between these and the phase curves of Barman et al (2005), calculated with a full
radiative transfer model for the no-redistribution case. The models show excellent agreement. This suggests
that, for reasonable models, the phase curve is determined primarily by the geometry and less by the
effects of changing temperature and atmospheric chemistry towards the limb. Indeed, we do not show any
phase curves indicating the effect of varying the $\gamma$ parameter, because the variation is quite
small ($\sim 7\%$ change in amplitude between $\gamma=10^{-2}$ and $\gamma=100$, with very little change in
shape).

\subsection{Temperature Profile}

We can also calculate the atmospheric temperature structure of these models, 
 using the mean intensity $J$ and
the condition of radiative equilibrium, $J = \sigma T^4/\pi$. 
\begin{equation}
T = \left( \frac{3}{4} \right)^{1/4} T_{eff}
\left[ \frac{2}{3} + \tau + \frac{4 \mu_0 I_0}{3 F_{int}} \left( 1 +
\frac{3}{2} \left( \frac{\mu_0}{\gamma}  \right)^2  - \frac{3}{2}
\left( \frac{\mu_0}{\gamma} \right)^3 \ln \left( 1 + \frac{\gamma}{\mu_0}\right)
- \frac{3}{4} \frac{\mu_0}{\gamma} e^{- \gamma \tau/\mu_0} \right) \right]^{1/4}, 
\end{equation}
where $F_{int} = \sigma T_{eff}^4$. Ideally, we would like to frame this using a quantity more
physically useful than $I_0$, so we shall define an equivalent temperature by $I_0 = \sigma T_{0}^4$.
Reframing the equation then, we have
\begin{equation}
T^4 = \frac{3}{4}  T^4_{eff}
\left[\tau + \frac{2}{3} \right] +  \mu_0 T^4_{0} \left[ 1 +
\frac{3}{2} \left( \frac{\mu_0}{\gamma}  \right)^2  - \frac{3}{2}
\left( \frac{\mu_0}{\gamma} \right)^3 \ln \left( 1 + \frac{\gamma}{\mu_0}\right)
- \frac{3}{4} \frac{\mu_0}{\gamma} e^{- \gamma \tau/\mu_0} \right]. \label{Tprof}
\end{equation}

Figure~\ref{Tplot1} and Figure~\ref{Tplot2} show the temperature profiles at a range of angles $\mu_0$,
for two cases ($\gamma=0.1$ and $\gamma=10$), and assuming $(T_{0}/T_{eff})^4=100$. For the case where energy is absorbed high
in the atmosphere the temperature profiles look pretty similar to the unirradiated case,
although the asymptotic value of the temperature as $\tau \rightarrow 0$ is higher the
closer one gets to the substellar point. In the case where energy is mostly absorbed
below the reradiative photosphere we find that the temperature profile has two regions
where temperature is almost constant -- the traditional one at low optical depth but
also a second one at $\tau>1$, where the bulk of the incoming flux is absorbed. This
is similar in character to the profiles found in many detailed radiative transfer models
(e.g., Seager, Whitney \& Sasselov 2000; Barman et al 2005, Burrows, Sudarsky \& Hubeny 2006).

The origin of this intermediate plateau is easily understood. The linearity of the equations
of radiative transfer and equilibrium means that the above solution corresponds to the superposition
of the solutions to two independent problems -- the standard solution radiating $F_{int}$ at the
boundary with no irradiation plus the solution corresponding to zero net energy transfer with 
the irradiation boundary condition imposed. As noted by Milne, the temperature profile arising
from this latter solution consists of a gentle transition between two asymptotic temperatures
(at high and low optical depth). The difference between the two asymptotes is driven by the
 last term in equation~(\ref{Tprof}), leading to a temperature jump 
$\Delta T / T_{0} = (3\mu_0^2/ 4 \gamma)^{1/4}$, 
 and so is larger for smaller $\gamma$. Thus, in
Figure~\ref{Tplot2} we see the behaviour of the zero flux solution at low optical depth
and then the transition to the normal thermal profile deeper inside the planet. The same 
behaviour is found in Figure~\ref{Tplot1} but the difference between the two asymptotic
temperatures is smaller (larger $\gamma$) and can barely be discerned in the figure.

\subsection{Spectral Lines}
\label{Spec1}

The presence or absence of spectral lines in secondary eclipse measurements is a topic
of recent excitement (Grillmair et al. 2007; Richardson et al. 2007; Swain et al. 2007). 
Although our model is founded on the grey atmosphere approximation, we can make simple
models of spectral line formation by considering a simple line formation model imposed on
the continuum structure represented by our model. 

Consider a spectral absorption line resulting from an excess opacity at some specific wavelength.
The line opacity at this wavelength is represented by $\tau' = (1 + \alpha ) \tau$. We may thus
determine the strength of the line using the absorption depth $A = 1 - F_{line}/F_{cont}$, where
\begin{equation}
F_{cont} = 2 \int_0^{\infty} S(\tau) E_2(\tau) d\tau
\end{equation}
is the continuum flux, $S$ is the standard source function, and 
\begin{equation}
F_{line} = 2 \int_0^{\infty} S_{line}(\tau') E_2(\tau') d\tau'
\end{equation}
is the flux at line center. The function $E_2 (x) = \int_1^{\infty} t^{-2} e^{-x t } dt$.

We assume local LTE, so that $S = B = T^4$, and we use the background temperature structure
from equation~(\ref{Tprof}). The resulting integration yields the expected
$F_{cont} = T_{eff}^4 + \mu_0 T_{0}^4$. The same calculation to determine $F_{line}$ requires
that we replace $\tau$ in equation~(\ref{Tprof}) with $\tau'/(1+\alpha)$ before evaluating
the integral. The resulting expression for $A$, in the limit $T_{eff} \ll T_{0}$, is
\begin{equation}
A = \frac{3}{2} \left( \frac{\mu_0}{\gamma} \right)^2
\left[ \alpha - \frac{\mu_0}{\gamma} \left( (1+\alpha)^2 \ln \left[ 1 + \frac{\gamma}{\mu_0 (1 + \alpha)} \right]
- \ln \left[ 1  + \frac{\gamma}{\mu_0} \right] \right) \right].  \label{Aline}
\end{equation}
In the case of $\gamma \rightarrow 0$ (deep absorption), we find
\begin{equation}
A \rightarrow \frac{1}{2} \frac{\alpha}{1+\alpha}.
\end{equation}
In the opposite limit of $\gamma \rightarrow \infty$ (high absorption), we find
\begin{equation}
A \rightarrow \frac{3}{2} \alpha \left(\frac{\mu_0}{\gamma}\right)^2 \rightarrow 0.
\end{equation}
Thus, we can expect strong absorption lines in the case where the optical opacity is much less
than the opacity in the infrared, but weak lines when the optical opacity is large. Similar
considerations apply in the case of emission lines ($\alpha<0$ and $A \rightarrow -A$).
Such behaviour is also consistent with the model proposed 
 in Richardson et al. (2007) for example, which invoke TiO or Silicate
clouds as sources which can absorb energy at low pressures (high in the atmosphere), in order
to explain the lack of observable water absorption.

\subsection{Transmission Spectroscopy}

Spectral line formation is also relevant to the question of absorption line spectroscopy during
primary transit (Charbonneau et al. 2002; Vidal-Madjar et al. 2003, 2004). However, it is somewhat
different in detail, since the observation is essentially measuring the attenuation of a parallel beam
illuminating and (partially) passing through the limb of the planet. Furthermore, we need to make
a further approximation, since the calculation of the transverse optical depth at the limb requires
a relation between the physical length scale and the optical depth scale.

We make use of the fact that the absorption occurs high in the atmosphere, where the temperature
profile is essentially isothermal. We note, however, that the $\tau \rightarrow 0$ asymptote is
still a function of angle from the substellar point. This can be seen from Figure~\ref{Tplot1} or
\ref{Tplot2}. Assuming that the equation of state is that of an ideal gas, we can approximate
the pressure as $P  = \rho g H$, where $H(\mu_0)$ is the local scale height. We can then use
hydrostatic equilibrium to derive a relation between density and optical depth, so that
$\rho = \tau/(\kappa H(\mu_0))$, where $\tau$ is still the same local radial optical depth 
as before. We can then also infer the relationship between optical depth and radial distance $r$
\begin{equation}
\tau = \tau_0 e^{-r/H(\mu_0)},
\end{equation}
where $\tau_0$ is a normalisation constant than can be chosen to fit this approximation (only
valid high in the atmosphere) to the true relation. Note that this relationship is a function
of angle, so that a surface of constant optical depth is not spherical.

Turning to the optical depth for absorption through the limb, which we denote as $\tau_{\ell}$, we
have
\begin{equation}
\tau_{\ell} = 2\gamma \int_0^{\infty} \kappa \rho dz  = 2\gamma \tau_0 \int_R^{\infty} \frac{ e^{-r/H(\mu_0)} r dr}
{H(\mu_0) \left( r^2 - R^2 \right)^{1/2}}
\end{equation}
where $R$ is the cylindrical radius and so $z^2+R^2 = r^2$ and $z = \mu_0 r$ (We also include a factor $\gamma$
to relate this optical absorption to the infrared opacity represented by $\kappa$).
 We can make the
evaluation of this integral easier by noting that the scale height is small compared to the radial
and cylindrical scales for the atmospheric base. As such, the exponential in the integrand will limit
significant contributions to scales such that $r - R \leq H_0$, where $H_0=H(\mu_0=0)$ is the 
atmospheric scale height at the limb. This comes about because the density drops rapidly as the 
light path moves away from $\mu_0=0$ and results in a negligible contribution to the absorption. With this
approximation the integral becomes
\begin{equation}
\tau_{\ell} = \gamma \tau_0 \left( \frac{2 R}{H_0} \right)^{1/2} e^{-R/H_0} \int_0^{\infty} y^{-1/2} e^{-y} dy
=\gamma \tau_0 \left( \frac{2 \pi R}{H_0}\right)^{1/2} e^{-R/H_0}. \label{TauTransmit}
\end{equation}
We can see then, that at cylindrical radii $R$, the optical depth to transmission is $\gamma (2 \pi R/H_0)^{1/2}$
larger than the radial optical depth at that location, meaning that the $\tau_{\ell} = 1$ surface will
lie higher in the atmosphere, where $\tau < 1$ (because $R \gg H_0$).

Using equation~(\ref{TauTransmit}), we can estimate the so-called `transit radius effect' (Baraffe et al. 2003,
Burrows et al. 2003), which is the difference between the radius measured by observing a planetary transit,
and that calculated in a traditional stellar model (which is usually determined at some fixed radial
optical depth $\tau_f$). Setting $\tau_{\ell} = 1$, to determine $R_T$, and setting $\tau=\tau_f$ to
determine $R_T-\Delta R$, we infer
\begin{equation}
\Delta R = \frac{H_0}{2} \ln \left[ \frac{2 \pi R_T}{H_0}\gamma \tau_f \right].
\end{equation}
Thus, the transit radius increment is proportional to the local scale height at the limb,
and only logarithmically dependant on $\gamma$ or $\tau_f$. Recall that $H_0 \propto T_{eff}/g$,
so that there is an inverse relationship with gravity. More surprisingly, there is no direct
dependance on the intensity of the illumination, since the asymptotic temperature at the
limb ($\mu_0 \rightarrow 0$) does not depend on $T_{0}$. Of course, the irradiation does
slow the planet cooling, so that $T_{eff}$ would be higher for the same planet under strong
illumination, but this is an indirect effect.

\section{Energy Redistribution}
\label{Finally}

The models in the previous section considered the absorption and re-emission of all radiation
locally, i.e. there was no energy transfer between different locations on the planet. However,
we have abundant evidence for strong atmospheric flows in our own giant planets and so it is
quite possible that winds and atmospheric circulation could redistribute the absorbed energy
over a significant fraction of the planet before it is reradiated. This would have a significant
influence on the observations, because the no redistribution model leads to large temperature
differences between the substellar and antistellar sides of the planet. 

\subsection{Simplest model}
\label{Redis0}

Thus, we turn now to the case of redistribution of energy over the surface of the planet.
In this section, we will return to the approximation of \S~\ref{Model1}, wherein we impose
the effects of irradiation by use of the outer boundary condition only.
 We do this in order
to isolate the physics of redistribution separately from any considerations about where
the energy is actually deposited. In the next section, we will combine our redistribution
model with the more detailed $\gamma$ models outlined in \S~\ref{Model2}.
Furthermore, we will assume the redistribution of energy over the surface happens
at a single, 
fixed optical depth $\tau_w$. This introduces a new source/sink term in
the radiative transfer equation to reflect the fact that energy may be added or
removed from the net flux, depending on location.
 Equation~(\ref{J0}) is thereby replaced by
\begin{equation}
\frac{1}{3} \frac{\partial^2 J}{\partial \tau^2} = - \Lambda \delta \left( \tau - \tau_w \right)
\end{equation}
where $\Lambda$ is a function (of position on the planet) describing the local energy input/loss due to 
atmospheric redistribution.

The solution to this equation is of the standard form $J = J_0 + J_1 \tau$ except that now
the first derivative experiences a discontinuous jump at $\tau = \tau_w$. Taking into account the
internal flux $F_{int}$ we find
\begin{eqnarray}
J & = &  J_0 + \left( \frac{3}{4} F_{int} + 3 \Lambda \right) \tau \qquad \left( \tau < \tau_w \right) \label{LowTau} \\
 & = &  J_0 + \frac{3}{4} F_{int} \tau \qquad \qquad ( \tau > \tau_w )
\end{eqnarray}
As before, we determine the constant $J_0$ by calculating first $I(0,\mu)$ and 
using that to calculate the net flux emerging from the surface
\begin{eqnarray}
F & = & 2 \int_{-1}^{1} \mu I(0,\mu) d\mu = 2 \int_{0}^{1} \mu I(0,\mu) d\mu
+ 2 \int_{-1}^{0} \mu I(0,\mu) d\mu \\
 & = & -  \mu_0 I_0 + 2 \int_{0}^{1} \mu I(0,\mu) d\mu
\end{eqnarray}
where we have again used as our upper boundary condition for incoming radiation a model in which
the incoming radiation is modelled as a beam of intensity $I_0$ coming from a direction $\mu_0$. 
Furthermore, this net flux has to equal the sum of $F_{int}$ and the effect of the $\Lambda$
term, whether it is positive or negative. Thus, $F = F_{int} + 4 \Lambda = F_{int} + \Delta F$. Using this to determine
$J_0$ results in a final answer for the specific intensity of
\begin{equation}
I(0,\mu) = \frac{3}{4} F_{int} \left( \mu + \frac{2}{3} \right) +  \mu_0 I_0
+ \frac{3}{4} \Delta F \left[ \mu (1 - e^{-\tau_w/\mu}) + \frac{2 - 3\tau_w e^{-\tau_w/\mu} }{3} 
+ 2 \left( e^{-\tau_w} -2 E_4(\tau_w)\right)\right]
\end{equation}
where $E_n(x) = \int_1^{\infty} e^{-x t} t^{-n} dt$ is the standard function found in radiative
transfer theory and we have made use of the identity $x E_{n-1} = e^{-x} - (n-1) E_n$. In the case of $\Lambda=0$, this reverts to equation~(\ref{Solution2}).

To model the redistribution, we need to specify the form of the function $\Lambda$.
We implement redistribution by imposing temperature equilibration on curves imposed
by the fluid circulation.
Recall that, in LTE, the specific intensity $J=B$, which is the local energy density, so that the redistribution of energy
essentially involves equilibration of the mean intensity $J$ around the appropriate curve. In particular,
using equation~(\ref{LowTau}),
we can calculate $J_w$, the mean intensity at the optical depth $\tau_w$ where redistribution
is assumed to occur
\begin{equation}
J_w = \frac{3}{4} F_{int} \left( \tau_w + \frac{2}{3} \right) + \mu_0 I_0 +
\frac{3}{4} \Delta F \left[ \tau_w + \frac{2}{3} + 2 \left( e^{-\tau_w} - 2 E_4(\tau_w)\right) \right]. \label{Jw}
\end{equation}
To complete the model, we also need to specify the geometry, i.e., along which curve 
is the temperature equilibrated? There are a variety of models for the fluid circulation on
the surfaces of irradiated planets (see \S~\ref{Hydro}), and each offers a different possibility.
We will base our model on the simulations of Cho et al (2003), who find that the strong winds
maintain a banded structure, with a hot equator and cold poles. This is qualitatively similar
to the bandedness of Jupiter and other solar system planets, although the bands are fewer
and wider because of the different rotational state of the system.
Thus, we enforce equilibration of $J_w$ on bands of constant $\theta_p$ (the angle measured
from the pole of the planet).
 Recall that  the angle cosine $\mu_0$ reflects the angle between the local normal
and the direction of the irradiating star. Casting this in terms of angles relative
to the pole, yields
 $\mu_0 = \sin \theta_p
\cos \phi_p$. The second term in equation~(\ref{Jw}) is linear in $\mu_0$ and so we wish
$\Lambda$ to have a similar dependance in order to cancel out some or all of that 
angular variation, in the interests of redistribution over $\phi_p$ for constant $\theta_p$.
Thus, we set $\Lambda = \beta_1 + \beta_2 \cos \phi_p$ on the substellar side, and
$\Lambda = \beta_1$ on the antistellar side (where we do not need the $\phi_p$ dependance to
cancel out any irradiation term). Energy conservation requires that, if we remove energy
at one point on the curve, we must inject it again somewhere else, so that
$\int_0^{2\pi} \Lambda d\phi_p = 0$, and so  $\pi \beta_1 + \beta_2 = 0.$. 
  Furthermore
total equilibration requires that all terms in equation~(\ref{Jw}) that depend on $\phi_p$
must cancel, yielding an expression for $\beta_1$
\begin{equation}
\beta_1 = \frac{ I_0 \sin \theta_p }{3 \pi \left( \tau_w + \frac{2}{3} + 2 e^{-\tau_w} - 4 E_4 (\tau_w) \right)}
\label{Complete}
\end{equation}
and the resulting mean intensity is 
\begin{equation}
J_w = \frac{3}{4} F_{int} \left( \tau_w + \frac{2}{3} \right)+ \frac{1}{\pi} I_0 \sin \theta_p.
\end{equation}
The intensity at the surface then follows,
\begin{equation}
I(0,\mu_0,\mu) = \frac{3}{4 } F_{int} \left( \mu + \frac{2}{3} \right) +
\frac{I_0 \sin \theta_p}{\pi} \left( 1 - \epsilon \right) + \epsilon I_0 \sin \theta_p \cos \phi_p,
\label{Icomplete}
\end{equation}
where
\begin{equation}
\epsilon = \frac{ \tau_w ( 1 + e^{-\tau_w/\mu} ) - \mu ( 1 - e^{-\tau_w/\mu})}
{ \tau_w + 2/3 + 2 ( e^{-\tau_w} - 2 E_4 (\tau_w) )}.
\end{equation}
This quantity $\epsilon$ 
encapsulates how the surface emission is affected by the depth at
which the redistribution occurs. In the limit $\tau_w \rightarrow \infty$, $\epsilon \rightarrow 1$,
in which case equation~(\ref{Icomplete}) becomes equation~(\ref{Solution2}). This is because, if
the redistribution occurs deep in the atmosphere, none of the irradiation actually reaches that depth, and the
absorbed energy is reradiated in situ as before. On the other hand, as $\tau_w \rightarrow 0$,
$\epsilon \rightarrow \tau_w/2 \rightarrow 0$, and the mean intensity distribution is independent of $\phi_p$,
although not $\theta_p$, reflecting our approximation that the temperature structure is now banded.

Let us now examine what fraction of the total luminosity is reradiated on the substellar side and the 
antistellar side.
Integrating $F = 2 \int_0^1 \mu I(0,\mu_0,\mu) d\mu$, and then integrating over $\theta_p$ and $\phi_p$,
we calculate
\begin{equation}
L_{sub} = \pi I_0 \left[ 1 - \frac{2/3 + E_4(\tau_w)}{\tau_w + 2/3 + 2 \left( e^{-\tau_w} - 2 E_4 (\tau_w) \right)} \right]
\end{equation}
and
\begin{equation}
L_{anti} = \pi I_0 \left[ \frac{2/3 + E_4(\tau_w)}{\tau_w + 2/3 + 2 \left( e^{-\tau_w} -2 E_4 (\tau_w) \right)} \right].
\end{equation}
 We see that $L_{anti} \rightarrow 0$
as $\tau_w \rightarrow \infty$, as expected. As $\tau_w \rightarrow 0$ on the other hand, $L_{anti}$ asymptotes
to almost $L_{sub}$, as expected in the case of complete redistribution. 
We will postpone calculation of most of the observables until \S~\ref{MoreApps}, but we will calculate the
expected full-phase flux, as this is the easiest way to estimate whether redistribution is taking place.
Recall that to calculate the full-phase flux, we calculate
 $F_{full} = 2 \int_0^1 \mu_0 I(0,\mu_0,\mu_0) d\mu_0$, i.e. we impose the condition
$\mu = \mu_0$. However, to do this easily, we need to change angular co-ordinate systems. The equilibration 
was performed on contours that encircle the pole of the planet, but the integration at full-phase
is most conveniently performed in angles measured from the substellar point. Thus, let us define
the angle $\phi_s$ as that which circulates about the substellar point, and $\theta_s$ the angle
relative to that point. As a result, $\sin \theta_p = ( 1 - \sin^2 \theta_s \sin^2 \phi_s )^{1/2}$.
The angle $\mu_0 = \cos \theta_s$ and we can thus perform the integral over $\theta_s$ and $\phi_s$.
 Figure~\ref{Fullf} shows the
resulting flux, expressed in terms of the effective $f_{eff}$ parameter of equation~(\ref{SupCon1}). 
We see that the value tends to the expected no-redistribution value of 2/3 as $\tau_w \rightarrow \infty$
and comes close to the expected full redistribution value of 0.25 as $\tau_w \rightarrow 0$. The
actual asymptotic value is
\begin{equation}
f_0 = \frac{1}{\pi^2} \int_0^{2 \pi} d\phi_s \int_0^1 d\mu_0 \mu_0 \left[ 1  - \left( 1 - \mu_0^2 \right)
\sin^2 \phi_s \right]^{1/2} = \frac{8}{3 \pi^2} = 0.2702.
\end{equation}

The resulting phase curves are shown in Figure~\ref{RedPhase}. For deep redistribution, the phase
curve is very similar to the no redistribution curve, while redistribution at low optical depth 
suggests almost complete uniformity.

\subsection{Redistribution in the $\gamma$-model}
\label{FinalModel}

We now wish to combine the effects of our models in \S~\ref{Model2} and \S~\ref{Redis0}, that is, allow for
both a different optical depth in the optical and IR and for redistribution. This will constitute
our final model. Thus, equation~(\ref{J0})
now has two source/sink terms
\begin{equation}
\frac{\partial^2 J}{\partial \tau^2} = -\frac{3}{4 } \gamma I_0 e^{-\gamma \tau/\mu_0} - 3 \Lambda
\delta ( \tau - \tau_w ).
\end{equation}

We follow the same approach as before to derive the mean intensity at $\tau_w$
\begin{eqnarray}
J_w & = & \frac{3}{4 } F_{int} \left( \tau_w + \frac{2}{3} \right) + 
\frac{3}{4 } \Delta F \left( \tau_w + \frac{2}{3} + 2 (e^{-\tau_w} - 2 E_4 (\tau_w)) \right) + \nonumber \\
&& + \mu_0 I_0 \left[ 1 - \frac{3}{4} \frac{\mu_0}{\gamma} e^{-\gamma \tau_w/\mu_0}
+ \frac{3}{2} \left( \frac{\mu_0}{\gamma} \right)^2 - \frac{3}{2} \left( \frac{\mu_0}{\gamma} \right)^3
\ln \left( 1 + \frac{\gamma}{\mu_0} \right) \right].
\end{eqnarray}

Once again, we wish to equilibrate this quantity on contours of constant $\theta_p$. The functional
dependance on $\phi_p$ is now more complicated than a single cosine, so let us denote the last
term as $ I_0 \sin \theta_p \Phi (\gamma, \cos \phi_p)$, so that all the variation is contained
in the function $\Phi$. Thus, equilibration requires
\begin{equation}
\Delta F = \beta_1 + \beta_2 \Phi 
\end{equation}
 to cancel out the variations on the
substellar side, while preserving the integral over the contour. Thus
\begin{eqnarray}
 \beta_1 & = & \frac{2}{3\pi} \, \, \frac{I_0 \sin \theta_p}{\tau_w + \frac{2}{3} + 2 ( e^{-\tau_w} - 2 E_4(\tau_w))}
\left( \frac{1}{\pi} \int_{-\pi/2}^{\pi/2} \Phi \, \, d\phi_p \right) \\
\beta_2 & = & -\frac{4}{3\pi} \, \, \frac{I_0 \sin \theta_p}{\tau_w + \frac{2}{3} + 2 ( e^{-\tau_w} - 2 E_4(\tau_w))}.
\end{eqnarray}
We will henceforth denote
\begin{equation}
< \Phi > = \left( \frac{1}{\pi} \int_{-\pi/2}^{\pi/2} \Phi \, \, d\phi_p \right).
\end{equation}
The resulting specific intensity at the surface is
\begin{equation}
I(0,\mu_0,\mu) = \frac{3}{4} F_{int} \left( \mu + \frac{2}{3} \right) + 
\frac{I_0 \sin \theta_p}{\pi} \epsilon \Phi +
\frac{I_0 \sin \theta_p}{2\pi} (1 - \epsilon) < \Phi > +
\frac{3 \mu_0  I_0}{4 } \left[ \frac{\mu_0}{\gamma} e^{-\gamma \tau_w/\mu_0} - \frac{\mu_0/\gamma}{1 + \gamma \mu/\mu_0} \right], \label{Ifull}
\end{equation}
where $\epsilon$ is defined as in equation~(\ref{Icomplete}). This looks much the same as the simpler version,
with an extra term on the end that accounts for the fact that the energy is deposited over a range
in optical depths, and some may be deposited above $\tau_w$ and some below.

We can now calculate 
 how much of the reradiated luminosity is emitted
from the substellar side, as a function of $\gamma$ and $\tau_w$. Figure~\ref{3Lg} shows 
the substellar and antistellar fractions as a function of $\tau_w$ for three different cases, $\gamma=100$,
$\gamma=1$ and $\gamma=0.01$.
At large $\gamma$, we see the expected behaviour, where large $\tau_w$ implies that
very little energy is reradiated on the antistellar side and small $\tau_w$ leads to almost
complete equilibration. For $\gamma \gg 1$, $\Phi \rightarrow \cos \phi_p$ and the last
term in equation~(\ref{Ifull}) is $\sim O(\gamma^{-1})$ even if $\tau_w \rightarrow 0$, so that we recover the same equations
as in \S~\ref{Model0}. This tells us that absorption high in the atmosphere can still result in
complete equilibration as long as $\tau_w$ is relatively small. However, absorption high in the atmosphere
results in little equilibration if $\tau_w > 1$.

In the case for $\gamma=1$ we see qualitatively similar behaviour,
except that the solution does not asymptote to complete equality as $\tau_w \rightarrow 0$.
The difference is even more extreme for $\gamma \ll 1$. Here
we see that there is an intermediate regime ($\tau_w \sim 3$) where almost complete equality
is established, but then that solution diverges again as $\tau_w \rightarrow 0$, with fully
three quarters of the luminosity emerging on the substellar side. This is because, in the limit
$\gamma \rightarrow 0$ (corresponding to deep deposition of energy), the limit of equation~(\ref{Ifull})
is
\begin{equation}
I(0,\mu_0,\mu,\gamma \rightarrow 0,\tau_w \rightarrow 0) = \frac{I_0 \sin \theta_p}{2 \pi^2}
+ \left( \frac{3}{4 \pi} I_0 \sin \theta_p \cos \phi_p \right) \mu,
\end{equation}
in which the second term now behaves like a traditional flux contribution, powered by irradiation energy
deposited well below the photosphere. The physical origin of this behaviour is that, in
this limit, many photons do not contribute to the temperature at low $\tau$ because they are
absorbed deep in the atmosphere (because the optical opacity is so low) and then escape from
below $\tau_w$ (which is small in this limit).

We can summarise the various effects by examining Figure~\ref{Lgt}, which shows contours indicating the fraction of
the total absorbed luminosity that is reradiated on the substellar side. We see that there are two
regimes which allow for approximate equality between substellar and antistellar sides. The first is
for $\gamma \gg 1$ and $\tau_w \ll 1$. This regime corresponds to the case when both absorption and redistribution
both occur high in the atmosphere, as might occur if one has a layer of stratospheric clouds and strong stratospheric winds.
 The second is the perhaps more traditional case of $\gamma \ll 1$ and $ 1 <\tau_w < 10$, where energy deposition
and energy redistribution occur at moderate (infra-red) optical depths.
Note also that the contours tend to change direction near the line $\gamma \sim \tau_w$. Above this
line much of the irradiation energy is absorbed in regions of the atmosphere above the redistribution
region, while below it the incoming photons penetrate to regions below $\tau_w$ before being 
absorbed.

\section{Some More Applications}
\label{MoreApps}

Let us now revisit the various observational probes with our new models and examine how
things are changed by the inclusion of redistribution.

\subsection{Secondary Eclipses}

We can, once again, calculate the full phase flux, as observed during secondary eclipse, using
equation~(\ref{Ifull})
 Once again, the flux integral is performed with $\mu=\mu_0$. The 
result is shown in Figure~\ref{fgt}, where we have again cast the result in terms of the $f_{eff}$
parameter. We see that, for $\tau_w \gg 1$, we recover the no-redistribution limit for all
$\gamma$ as expected. For a broad range of low and intermediate values of $\tau_w$, the value of $f_{eff}$ varies
between 0.4 and 0.5.
For low $\gamma$, the value of $f_{eff}$ actually has a minimum near $\tau_w \sim 1$. 
For large $\gamma$, the behaviour is more like the simple model, with near complete
redistribution at low $\tau_w$, changing monotonically to no redistribution at large $\tau_w$.

The diversity of values obtained demonstrates the complexity of inferring planetary properties from
as simple an observation as comparing a flux level at a single phase to the models. We note that the
value $f_{eff} = 0.25$ that one expects from the naive full redistribution model is realised only in
the $\tau_w \ll 1$, $\gamma \gg 1$ limit and that $f_{eff}$ attains intermediate values $\sim 0.4$--$0.6$
over most of the parameter space. Thus, while we may be able to demonstrate, using this kind of observation,
that a planet possesses some level of redistribution, it will be difficult to determine specific parameters
without additional information.

\subsection{Phase Curves}

This behaviour is also reflected in the diversity of phase curves, shown in Figure~\ref{Phases}.
The three panels show the phase curves (for edge-on orbits) for three choices of $\gamma$. In
each panel we show curves for three different $\tau_w$. As expected, large $\tau_w$ makes for
a strong variation with phase, while low $\tau_w$ results in a very flat phase curve. At low
$\gamma$ values, this distinction is blurred somewhat.

Figure~\ref{fphase} summarises this behaviour. We show here the amplitude of the observable
phase variation $\delta f$, from peak to trough, expressed once again in terms of $f_{eff}$. We see,
not surprisingly, similar behaviour to Figure~\ref{fgt}. In the large $\tau_w$ limit, we
recover $\delta f > 0.6$, essentially no redistribution. On the other hand, amplitudes
of $\delta f<0.1$ are obtainable in the same two regimes as identified above. Once again, intermediate
behaviour is recovered over most of the parameter space, with phase variations $\sim 50\%$ of the
no-redistribution model expected.

\subsection{Limb Darkening}

Figure~\ref{RedLimb} shows the influence of redistribution on the limb-darkening of the planet,
when observed at full phase. With the model
of redistribution adopted here, the limb is no longer at a constant intensity, but rather
shows a variation $\propto \sin \theta_p$. Figure~\ref{RedLimb} shows contours of the contrast observed
at full phase between the intensity at $\theta_p=\pi/2, \phi_p=\pi/2$ (the limb at the equator)
and the intensity at the center $\theta_p=\pi/2, \phi_p=0$. We see that deep redistribution ($\tau_w \gg 1$)
 yields low values, indicating 
strongly darkened limbs, as in the no-redistribution case. We also find dark
limbs in the limit $\gamma \ll 1$ and $\tau_w \ll 1$, where the energy is absorbed deep in the
star but redistribution occurs high in the atmosphere. However, for $\gamma \gg 1$ (absorption
high in the atmosphere), we actually
find equality and possibly even limb-brightening. Similar effects have been seen in more detailed
models, for instance when the addition of metals results in a large amount of stratospheric
absorption (Fortney et al 2006).

\subsection{Temperature Profiles}

We can also examine the temperature profile.
For $\tau<\tau_w$, 
\begin{eqnarray}
T^4 & = &  \frac{3}{4}  T^4_{eff} \left( \tau + \frac{2}{3} \right) + T_{0}^4 \sin \theta_p \left[ \left( \left[ 1 + \frac{\tau - \tau_w}{\tau_w + 2/3 + 2 (e^{-\tau_w} - 2 E_4 (\tau_w))}\right]
 \frac{< \Phi >}{2} + \right. \right. \nonumber \\
 && \left. \left.  \frac{\tau - \tau_w}{\tau_w + 2/3 + 2 (e^{-\tau_w} - 2 E_4 (\tau_w))} \Phi 
+ \frac{3}{4} \frac{\mu_0 \cos \phi_p}{\gamma} \left[ e^{-\gamma \tau_w/\mu_0} - e^{-\gamma \tau/\mu_0} \right] \right) \right]. \label{Tprofw}
\end{eqnarray}
For $\tau >\tau_w$, the temperature profile still obeys equation~(\ref{Tprof}).

Figure~\ref{Tprof1} shows temperature profiles for two cases where $\gamma=100$. In the upper panel,
we show the result for shallow redistribution ($\tau_w=0.1$) and in the lower panel, deep redistribution
($\tau_w=10$). In each case, the solid curve is the temperature profile at the substellar point ($\theta_p = \pi/2$, $\phi_p = 0$) and
the dotted curve is for the limb ($\theta_p=\pi/2$, $\phi_p = \pi/2$). In each case, the ratio
$ (T_{0}/T_{eff})^4 = 100$ and the temperature is measured relative to the true internal $T_{eff}$.
In the case of $\tau_w=10$, the effect of the temperature jumps are largely erased by the time
$\tau \rightarrow 0$, presumably because most of the absorbed energy never reaches $\tau=10$ when
$\gamma=100$. In this case, a temperature inversion is evident, and
 an important part of the energy balance at moderate $\tau$ is thus the
transport of energy inwards by thermal radiation from overlying layers heated by the stellar photons.
Thus, equilibration at $\tau_w=0.1$ does a much better job of smoothing out the temperature
variations.

In the opposite limit ($\gamma=0.01$), deep absorption, the temperature profiles are shown in
Figure~\ref{Tprof2}. In this case, the opposite behaviour is seen. The deep redistribution
case is effective, because most of the irradiation energy is deposited in deeper layers
and is included in the thermal balance at $\tau=\tau_w$. In the shallow redistribution case, 
the temperatures near the substellar point are much hotter at $\tau \sim 1$ and so this case
shows a strong asymmetry. There is an interesting center-to-limb temperature inversion at
low $\tau$ for this case, but it will be difficult to probe with any vertically integrated measures.

\subsection{Spectral Lines}

We can evaluate the integrals for spectral line formation in a similar fashion as in \S~\ref{Spec1},
although the algebra is somewhat more painful this time, and so we will not write the full
expression down here. Figure~\ref{Line1}
shows the contours of $A=1 - F_{line}/F_{cont}$ that result from the appropriate integration
 in the case where $\alpha=0.3$, for the usual
range of $\gamma$ and $\tau_w$. The calculation is once again at full phase orientation. We see that
 the strength of the line is relatively weak for $\gamma>1$ and rather insensitive to $\tau_w$. This is
what we expect on the basis of Figure~\ref{Tprof1}, because the temperature profile is close to isothermal.
For $\gamma <1$, the results are more sensitive to $\tau_w$, and can be quite strong in the case where
$\tau_w$ has a value only slightly larger than unity. This is because the combination of sharp temperature
jumps and different opacity scales can lead to large contrasts between line and continuum.

\section{Discussion}
\label{Finis}

The overall goal of this exercise is to develop a simple framework within which one can examine
the qualitative features of the irradiated atmosphere problem. Clearly, a simple, semi-analytic model
such as this cannot replace numerical models that incorporate detailed treatments of radiative transfer,
chemical equilibrium (and disequilibrium) and hydrodynamics. However, the complexity of the various
physical inputs into these models often results in a situation in which the physical understanding lags
behind the calculational output. As a result, simplified models provide a pedagogical framework which
can guide the more detailed models. This is likely to be particularly useful as we try to develop physical
models for the redistribution of energy within an irradiated planet atmosphere. This is likely to be a 
difficult and costly enterprise, and the simplified models may help to illuminate the nature of the desired
solution and possible pitfalls on the path to their realisation.
As a first step, we wish to cast our models in terms of the more detailed models in the literature, in
the sense of describing which of our simple models correspond to the more detailed models.

The first attempts to treat the structure of irradiated giant planets (Saumon et al. 1996; Guillot et al. 1996)
used essentially grey atmosphere models with a simple asymptotic temperature criterion. The first real
attempt to model the radiative transfer under conditions appropriate to Hot Jupiters was by Seager \& Sasselov (1998). These models were dust free, and the incoming optical/UV radiation was found to penetrate well below the
infra-red photosphere. These models therefore fall into the $\gamma <1$ category, and establish a theme
for most of the subsequent literature. This was followed by Barman, Hauschildt \& Allard (2001), who produced
two sets of models, ``clear'', in which condensates were removed by a rainout process, or ``cloudy'', in which
the condensates were assumed to form clouds in situ. The cloudy models were found to yield clouds extending from the
photosphere upwards to low optical depths, and most likely correspond to the $\gamma >1$ case. This is supported
by the claim that this led to much smoother emergent spectra than in the clear case, as expected for spectral
features in the $\gamma \gg 1 $ model.

Sudarsky, Burrows \& Hubeny (2003) present a variety of models, covering a large range of irradiations. The
models they classify as classes IV and V are most appropriate to Hot Jupiters. The majority of their models
clearly fall within the $\gamma <1$ regime. This can best be illustrated by the temperature profiles.
Their Figure~12 shows how the temperature profile of a planet changes as it is moved closer to a G0V primary.
At distances $<1$~AU, the temperature profile developes a plateau at pressures $>$1~bar, evolving to a shape
very similar to that in Figure~\ref{Tplot1}. For the hottest planets, Iron and Forsterite condensates are
possible at high altitude, leading to smaller plateaus, as might be expected for larger $\gamma$ atmospheres.
The clouds in this case are localised, however, and so the analogy with the $\gamma$ models is not as good
as for the Barman et al. (2001) models.

For the purposes of calibration the
models of Barman, Hauschildt \& Allard (2005) are 
 of particular interest because they
   calculated a truly two-dimensional
model and so produce
temperature profiles as a function of angle from the substellar point, rather than just a single `representative'
profile. We can thus accurately calibrate
the effective $\gamma$ of their model by making use of the temperature value of the observed plateau in the
model. We calibrate $T_{0}$ using the known properties of the HD~209458b system, $T_{0} = T_{eff} (R_*/a)^{1/2}=2030$~K. 
 Taking the profile from their Figure~4 for HD~209458b at the substellar point, the high temperature
plateau has a value $\sim 2500$~K. We identify this value with the $\tau \rightarrow \infty$ limit of the
irradiation term in 
 equation~(\ref{Tprof}). We use this expression for clarity since the
Barman et al. models are for the case of no redistribution, but equation~(\ref{Tprofw}) reduces to this in
the limit of $\tau_w \rightarrow \infty$. The temperature of this plateau is
\begin{equation}
T^4 = T_{0}^4 \left[ 1 + \frac{3}{2}\left( \frac{\mu_0}{\gamma}\right)^2 - \frac{3}{2}
\left( \frac{\mu_0}{\gamma}\right)^3 \ln \left( 1 + \frac{\gamma}{\mu_0} \right)  \right] = 2500^4. \label{Trav2}
\end{equation}
We can solve this to infer $\gamma \sim 0.45$ for $\mu_0=1$. Figure~\ref{Func} shows the expression 
\begin{equation}
\Upsilon = 1 + \frac{3}{2}\left( \frac{\mu_0}{\gamma}\right)^2 - \frac{3}{2}
\left( \frac{\mu_0}{\gamma}\right)^3 \ln \left( 1 + \frac{\gamma}{\mu_0} \right)  
\end{equation}
as a function of $\gamma$. We see that the inner plateau temperature increases as $\gamma$ decreases, with
the energy getting deposited at greater depth. The lower asymptote for these values of $\gamma$ and $T_{0}$
is derived from $\Upsilon - 3\mu_0/4\gamma$ (shown as the lower curve in Figure~\ref{Func}), and yields
a temperature $\sim 1800$~K. Furthermore, Barman et al. use an internal luminosity effective temperature
of $T_{eff}=230$~K. The resulting temperature profiles are shown in Figure~\ref{BHA}. The agreement with
Figure~4 of Barman et al is excellent for
$\tau > 1$ (corresponding to $P > 10$~mbar). At low optical depths, the temperature drops lower in the
detailed models, although it does eventually asymptote to higher temperatures off the edge of the
given plot (Barman, private communication). This behaviour presumably reflects the non-monotonic nature
of the real opacity at high altitudes. A general feature of these kinds of model is that (infrared) optical
depth unity lies somewhere in the middle of the transition between the two plateau. This can be seen
in other models as well, such as from the location of the `brightness temperature region' in Figure~1 of Fortney et al. (2005).

\subsection{Redistribution as dilution}

The detection of secondary eclipses in the HD209458 and TrES-1 systems generated a flurry of theoretical
activity. Of particular interest was the question of whether energy was transferred from the day side to
the night side. However, all of these models were flux-conserving and so approximate redistribution by
diluting the incoming flux by an appropriate factor $f$.
 Burrows,
Sudarsky \& Hubeny (2005) compare their $f=0.25$ (full redistribution) models to the observed data and
find that they are too faint, suggesting that less dilution is best. BSH also express surprise
 that adding iron or forsterite clouds to their models does not change the predicted fluxes by an appreciable
amount. We can understand this by noting that increasing $\gamma$ at fixed $\tau_w \sim 100$ does not change
the predicted $f_{eff}$, as was noted in \S~\ref{Sec1} and can be seen from Figure~\ref{fgt}. Fortney et al. (2005)
find that they cannot fit the two IRAC measurements for TrES-1 with a model without redistribution, 
but are able to almost fit the data with a diluted radiation field appropriate to significant
redistribution. The temperature profiles in their Figure~1 suggest again that their model lies in the
$\gamma < 1$ regime, with the characteristic two-plateau behaviour very evident. They note that the inclusion
of an arbitrary extra continuum opacity in the range 2--6$\mu$m would improve the fit, and which would also move
their model to larger $\gamma$. Seager et al. (2005) include, as an additional constraint, a limit on the flux
variation at wavelengths $\sim 2.2\mu$m (Richardson et al. 2003). This places limits on the depth of observable
water bands at near-IR wavelengths. Combining this with the secondary eclipse measurements, they found that
the clear ($\gamma < 1$) models were ruled out, but models with thick, absorbative clouds did match the models.
In this case clouds composed of Silicates extend from 0.8--5~mbar and Iron condensates from 1.3--10~mbar. From
their Figure~4, we can infer directly the approximate $\gamma$ values appropriate to their models. In the case
of the cloud-free models, the optical photosphere lies at greater pressures than the infrared photosphere, and
$\gamma \sim 0.1$, while the cloudy model has $\gamma \sim 10$. Similarly Barman et al. (2005) find that their
no-redistribution model overpredicts the flux at IRAC wavelengths, so they too infer that some level of energy
redistribution must be occuring.

Subsequently, Fortney et al. (2006) have used enhanced metallicity models to predict the structure of
the newly discovered hot Jupiter HD149026~b. Their models with 3 times solar metallicity, when calculated
in the `4$\pi$' (full redistribution) case, have little cloud absorption because the relevant condensation
curves lie at pressures of hundreds of bars. In the `2$\pi$' (no redistribution) case, the models result
in hot stratospheres because the TiO and VO absorption is present at lower pressures for such hot models.
The comparison between the resulting temperature profiles can also be understood in the context of our
models, since the increase in $\gamma$ for the no redistribution model leads to a smaller temperature
difference between the two plateau, as observed. In this case though, even the hotter model still lies
in the $\gamma <1$ regime.

Burrows, Sudarsky \& Hubeny (2006) make a first attempt at predicting phase curves while
including a primitive redistribution prescription, in which they remove a fraction $P_n$ of the
irradiation energy (i.e. this is still a dilution model) and then incorporate it as an extra internal energy source on the antistellar side. Their
temperature profiles again very clearly conform to the $\gamma <1$ structure we have seen before,
consistent with their neglect of clouds. They do, however, discuss one `extreme' model with significant
heating of the stratosphere using significant TiO and VO absorption in the upper atmosphere. This may
indeed lie with the $\gamma >1$ regime. Burrows et al present phase curves for a variety of values of
$P_n$ and bandpasses, and note that even significant redistribution can still leave some remnant phase
variation because of the non-uniform temperature pattern on the substellar side. This has some qualitative
resemblance to the kinds of variation seen in Figure~\ref{fphase}. 

\subsection{Including redistribution in the radiation field directly}

However, all of the previous attempts to model redistribution have relied on some kind of dilution
of the original radiation, rather than removal of the energy at the appropriate point in the atmosphere.
The models discussed here allow us to examine how the temperature profile changes if we actually include
sources and sinks at different points in the atmosphere.

Let us consider the effect of redistribution at different $\tau_w$, using our calibrated
version of the Barman et al. (2005) model, shown in Figure~\ref{BHA}. Figure~\ref{BH3} shows the effect,
on the temperature profile at the substellar point, 
 of moving
$\tau_w$ to progressively smaller values. 
Because this is for the substellar point, redistribution removes energy at the specified location,
resulting in some pronounced local temperature inversions. In fact, for $\tau_w \sim 10^2$--$10^4$,
there is a pronounced local peak in the temperature, resulting from a competition between the local
deposition of irradiation energy and the removal by redistribution. At low enough $\tau_w$, we start
to see a significant drop in the stratospheric temperature. The equatorial variation in temperature
profile, from center to limb, is shown in Figure~\ref{BH6}, for the case $\tau_w = 100$. We see that
the effect of redistribution changes from a sink to a source for $\mu_0<0.4$, although the general
character of a hot substellar point is retained, albeit with reduced amplitude. In the case where
$\tau_w = 1$, shown in Figure~\ref{BH7}, the temperature at high altitudes is much more uniform.

These effects apply for the case of $\gamma < 1$, which applies to most of the models in the
literature. However, there is tentative evidence that some systems may favour the $\gamma >1$ regime,
so let us consider redistribution in this case. First, we show in Figure~\ref{BH2} the effect of changing $\gamma$,
while keeping all other inputs fixed. The dotted line shows our calibrated model for $\gamma=0.45$. Lower values
of $\gamma$ result in even higher temperature plateau, as expected, while larger $\gamma$ drives the atmosphere
closer to an isothermal state. This is precisely why the observed lack of spectral lines favours a high $\gamma$ model.
Figure~\ref{BH4} then shows the effects of different $\tau_w$ at the substellar point, but this time for
the $\gamma=10$ model. The temperature jumps are less in this case, but there are still temperature
inversions directly above the layer where redistribution occurs. However, the stratospheric temperatures do
return to close to their value in the $\tau_w \rightarrow \infty$ case,  with large differences only becoming
apparent at $\tau_w < 5$. We note, however, that it is precisely this region (large $\gamma$, moderate $\tau_w$)
that seems to be preferred by the data for HD209458b and HD189733b.

\subsection{Hydrodynamic redistribution models}
\label{Hydro}

We have thus far discussed the attempts to model energy redistribution in terms of calculations
of radiative transfer. A second, and mostly independent, body of work is focussed on the hydrodynamics
of irradiated atmospheres and how this might, in turn, influence the redistribution of energy.

The complexity of the physics involved has resulted in somewhat divergent treatments of the phenomenon,
each placing emphasis on different aspects of the phenomenon. Showman \& Guillot (2002) performed three-dimensional
fluid dynamical simulations of the atmosphere, finding the existence of broad east-west jets and strong 
residual temperature contrasts at the photosphere. Their models also suggested that some fraction of the
irradiation energy might get transported deeper into the planet, where it might help to explain the
anomalously large radius of HD~209458b. Iro et al. (2005), however, found that very little energy was
deposited at pressures $>100$~bars. Iro et al. use a coarse version of a one-dimensional radiative transfer
model, and include temporally variable forcing to mimic the planetary rotation. However, they do confirm
the existence of a strong phase variation within the context of their model, which also fits into the $\gamma <1$
class of our models. In an updated version of their three-dimensional model, Cooper \& Showman (2005) quantify
their phase variation, predicting that the equatorial fluid motions should shift the peak of the
phase variation $\sim 60^{\circ}$ downstream, although subsequent revisions of the model quoted in Knutson et al. (2007b) find a smaller phase shift. Similar behaviour is found by Langton \& Laughlin (2007).
An entirely different approach was taken by Cho et al. (2003; 2008), using a high resolution, two-dimensional model of the fluid
equations on a rotating sphere, treated in the shallow water approximation. Cho et al. find  generic behaviour
in which the principal hydrodynamical feature is a small number of broad equatorial bands, accompanied by coherent
polar vortices that rotate about the pole on timescales of several planetary rotation periods. The motion of
the temperature field is found to be quite diverse, depending on the strength of the heating and momentum transport
in the atmosphere and where the energy is deposited. 

Fluid motions are responsible not only for equilibration, but potentially also mixing. Cooper \& Showman (2006)
showed that hydrodynamic mixing processes can smooth out temperature and compositional differences between different
parts of the atmosphere. They predict that this will lead to homogeneous Carbon chemistry, with most of the
Carbon residing in CO and very little in CH$_4$ at low ($<$1 bar) pressures. Furthermore, the resulting temperature
structure is considerably more isothermal and model spectra show weak features (Fortney et al. 2006), in
similar fashion to the observations. It is interesting to consider these latter models in the light of our $\gamma$
parameter. A uniform vertical redistribution of energy by hydrodynamic mixing can be considered equivalent to
uniform absorption throughout the atmosphere, and can lead to a higher effective $\gamma$ than might be inferred
simply from the opacity. 

Examination of Fortney et al. (2006) suggests another application of our method. They use a global dynamical
simulation of the atmosphere to calculate how the temperature profile varies across the face of the planet
and then calculate many one-dimensional radiative transfer models to quantify the likely spectral appearance
as determined by the true two-dimensional surface variations. However, they were not able to iterate between
the dynamical model and radiative transfer model to achieve a true radiative equilibrium because of the prohibitive computational expense that would entail. A model such as the one described here, suitably parameterised by
an effective $\gamma$, could be used as an intermediate approximation to enforce a global energetic equilibrium
while still retaining the spirit of the radiative transfer model. The resulting temperature distribution could
then be used to calculate the detailed appearance.

\section{Conclusion}

In summary, the 
 intent of the models presented here was to understand the qualitative behaviour of 
radiative transfer in hot Jupiter atmospheres. The simplified nature of the models is
dictated by the difficulties of including both full radiative transfer and detailed hydrodynamics,
both of which are important parts of a proper understanding of the phenomenon.
 In particular we have studied how the models
change depending on the parameters $\gamma$ (which reflects the difference in opacity for
incoming optical-wavelength radiation and outgoing infrared-wavelength radiation) and
$\tau_w$ (the depth at which energy is redistributed horizontally across the planets atmosphere).

We have also studied how the various new observational probes of Hot Jupiter atmospheres are
influenced by changes in these parameters. We find that models in which redistribution occurs
at large infra-red optical depths redistribute little energy to the night side (as expected).
Perhaps more surprisingly, we find that significant day/night flux differences can also be found
in the case $\tau_w<1$, as long as $\gamma < 1$ as well. We find that day-side temperature inversions
are a generic feature of atmospheres in which the redistribution occurs at moderate ($\tau_w \sim 1$--1000)
optical depths. This is of interest because of recent claims that temperature inversions have
been found in HD~209458b (Knutson et al. 2007). The initial explanation for such behaviour is
high altitude absorption (Burrows et al. 2007; Fortney et al. 2007), although the very low albedo (Rowe et al. 2007)
restricts the absorbing species to be a poor reflector. Some level of redistribution is required to explain
the HD~209458~b secondary eclipse amplitude and (lack of) phase variations, so that another potential
explanation for the temperature inversion may simply be that it is related to the depth at which the
energy is removed from the day-side. Certainly, figures~\ref{BH6} and \ref{BH7} show that some parts of
the planet experience pronounced temperature inversions, even without a significant high altitude absorber.
However, it will require detailed radiative transfer models to confirm whether this is a viable model
or not.

Our initial exploration has  been far from exhaustive. One important advantage of these toy grey atmosphere models
is their flexibility, and we have studied how energy redistribution changes the character
of the models. This will hopefully provide a guide as to how one might perform similar
calculations using the more detailed models, which presently can only consider changes at either
the top or bottom of the atmosphere.
 Furthermore, we have also not exhausted the parameter space of toy models.  For instance, one can imagine many
other redistribution schemes apart from the one we have implemented here.

\acknowledgements This work has been supported in part by NASA ATP contract NNG04GK53G and by
funds associated with Spitzer Space Telescope program GO-20101. The author thanks Sara Seager
and Travis Barman for helpful comments.

\newpage

\plotone{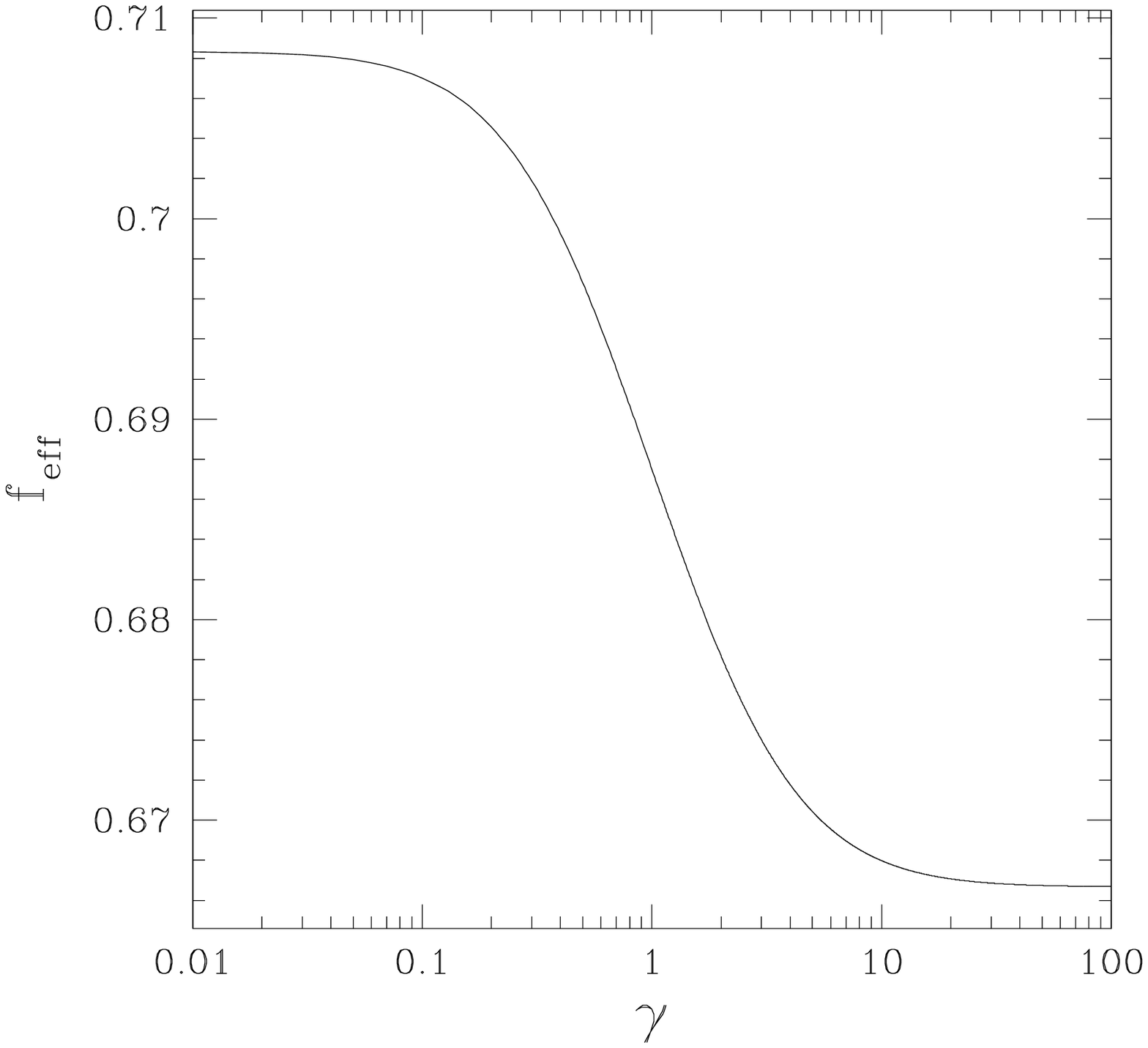}
\figcaption[f1.ps]{ The flux observed at superior conjunction, as a function of the energy deposition
depth parameter $\gamma$, and expressed as an effective value of the parameter $f$. We see that
the valeue of $f_{eff}$ is rather insensitive to $\gamma$.\label{Opp}}
\clearpage
\plotone{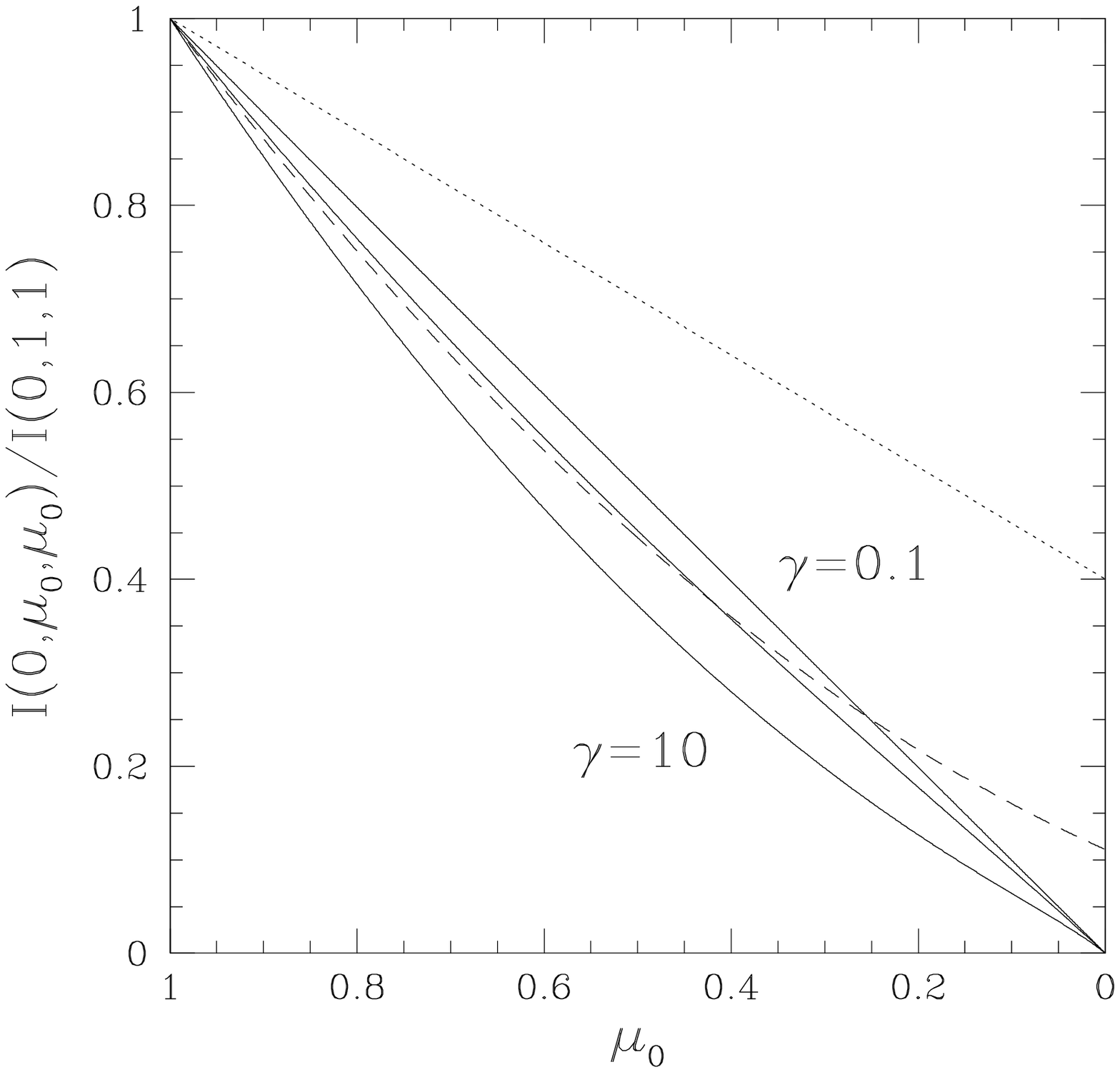}
\figcaption[f2.ps]{ The dotted line indicates the normal limb darkening law for an
unirradiated grey atmosphere. The three solid lines indicate the same for  our irradiated no-redistribution
models, for three different values of the deposition depth parameter $\gamma$. The values are (from the left) $\gamma=10$, $\gamma=1$ and $\gamma=0.1$. The dashed line represents the expression in Milne (1926),
equation~14. \label{Limb}}
\clearpage
\plotone{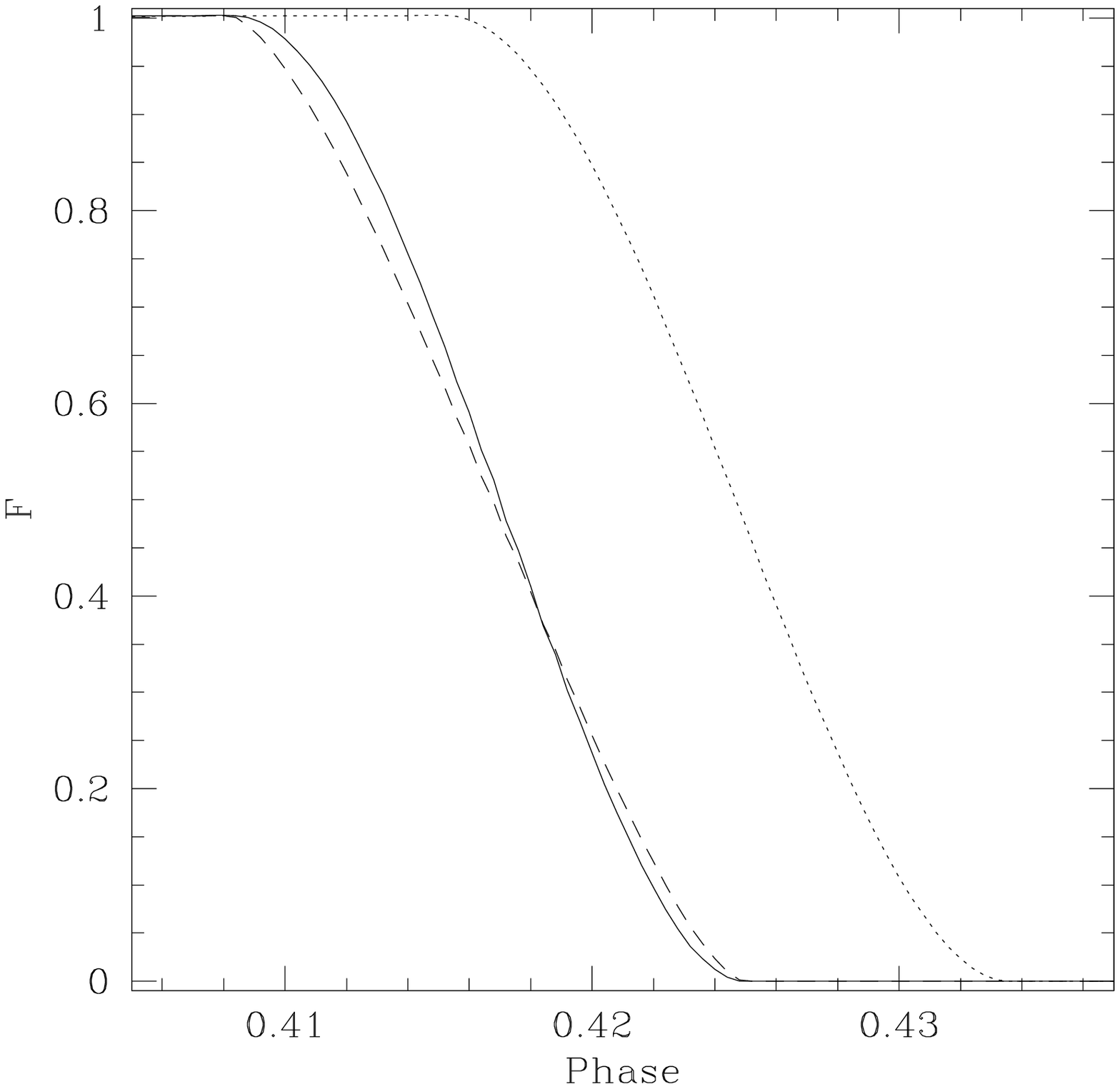}
\figcaption[f3.ps]{ The solid and dotted curves represent the secondary eclipse ingress for the
no-redistribution model, for zero and 2$^{\circ}$ inclination respectively. The dashed line is the
edge-on ingress for a uniform hemisphere model. \label{Trans}}
\clearpage
\plotone{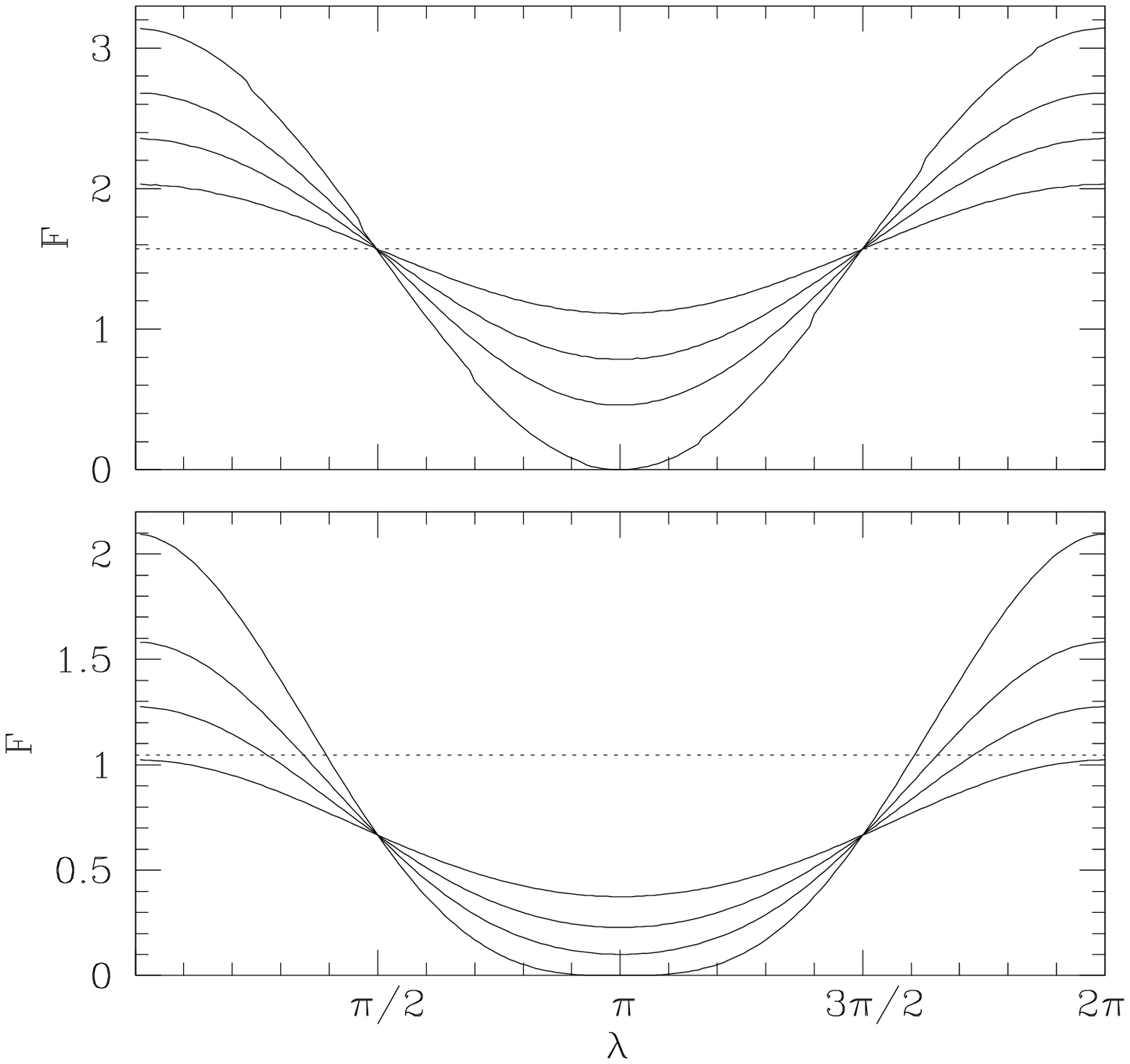}
\figcaption[f4.ps]{ The upper panel shows phase curves for the case of the uniformly bright hemisphere model.
The lower panel shows phase curves for the case of the no-redistribution model. Curves are shown in each case
for inclinations of $0^{\circ}$,$45^{\circ}$,$60^{\circ}$ and
$75^{\circ}$. The edge-on cases are given by the analytic solutions in equations~(\ref{FlatSol}) and (\ref{SpotSol}) respectively.
\label{Phase12}}
\clearpage
\plotone{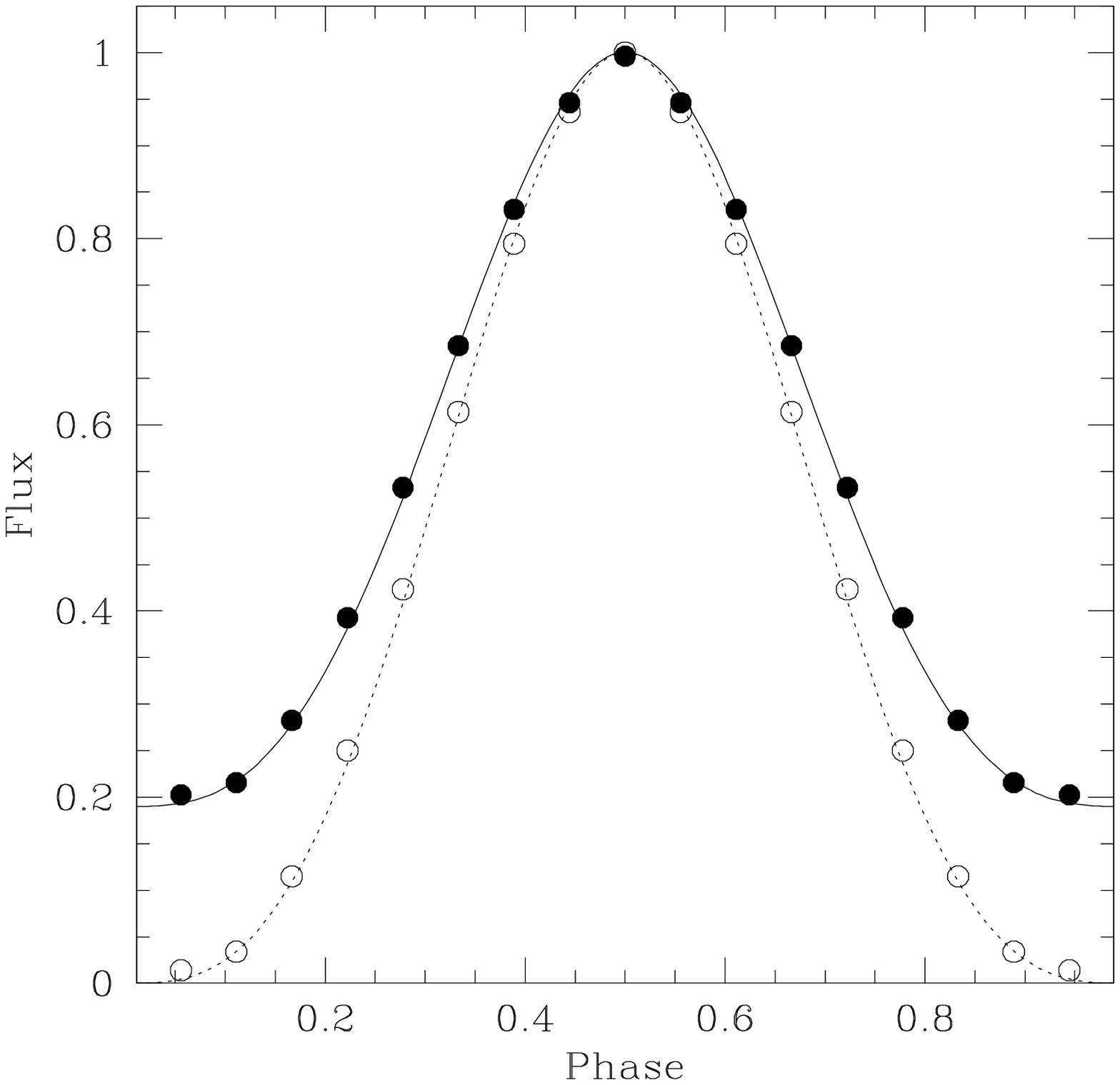}
\figcaption[f5.ps]{The solid and open circles represent phase curves for $24 \mu$m and $8\mu$m respectively,
calculated using the models of Barman, Hauschildt \& Allard (2005). The dotted line represents the no-redistribution model described in equation~(\ref{SpotSol}). The solid line is the same solution, but superimposed
on a DC level of 0.19 on this scale, to represent a contribution from internal luminosity. The excellent
agreement suggests that the phase curves are determined primarily by geometrical effects. \label{Phase3}}
\clearpage
\plotone{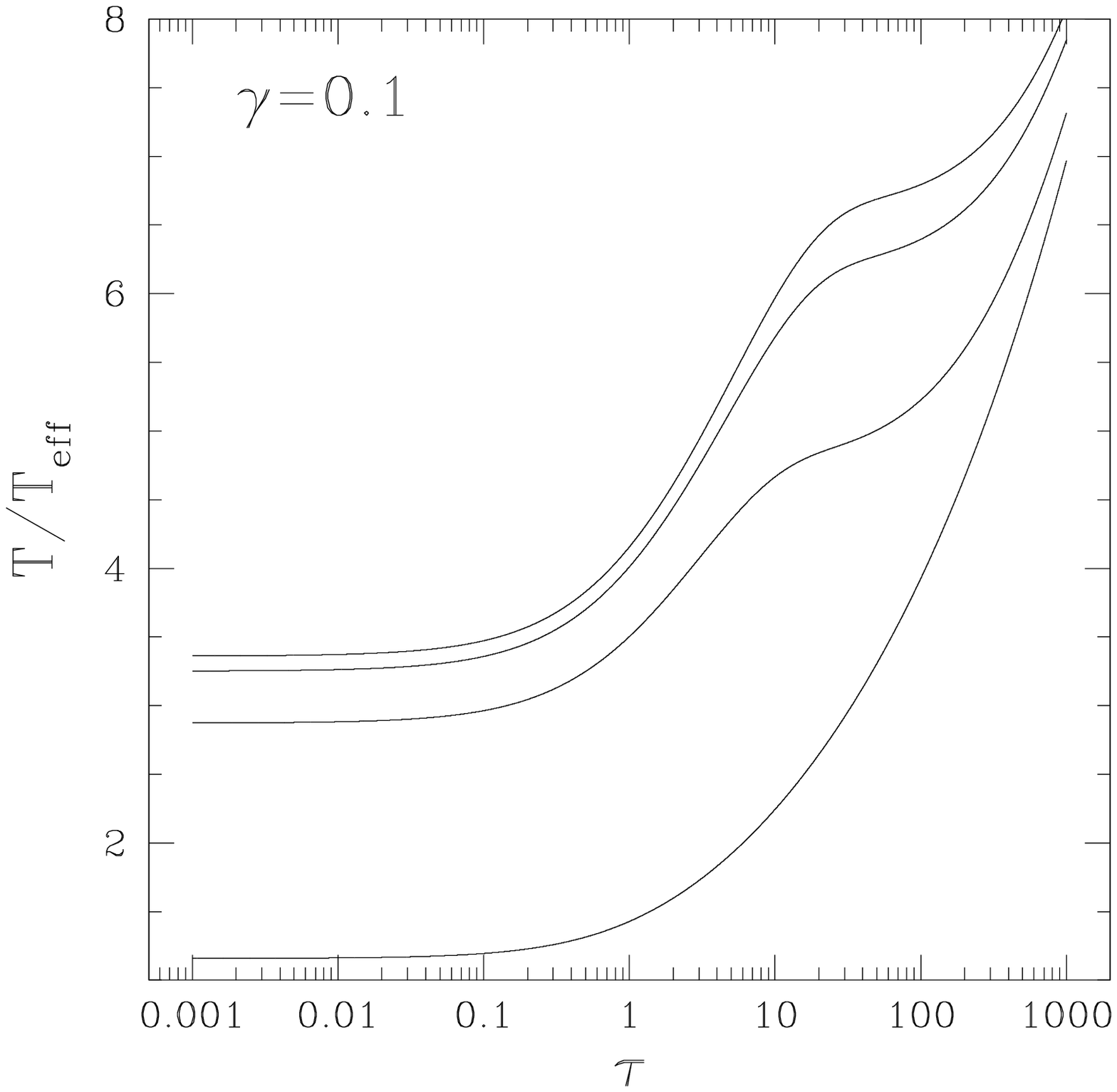}
\figcaption[f6.ps]{ The temperature profiles for angles of 0,30,60 and 90$^{\circ}$ to
the line of sight, for the case $\gamma=0.1$, i.e. where the irradiation penetrates to below
the reradiation photosphere. We have also used $(T_{0}/T_{eff})^4 = 100$.\label{Tplot1}}
\clearpage
\plotone{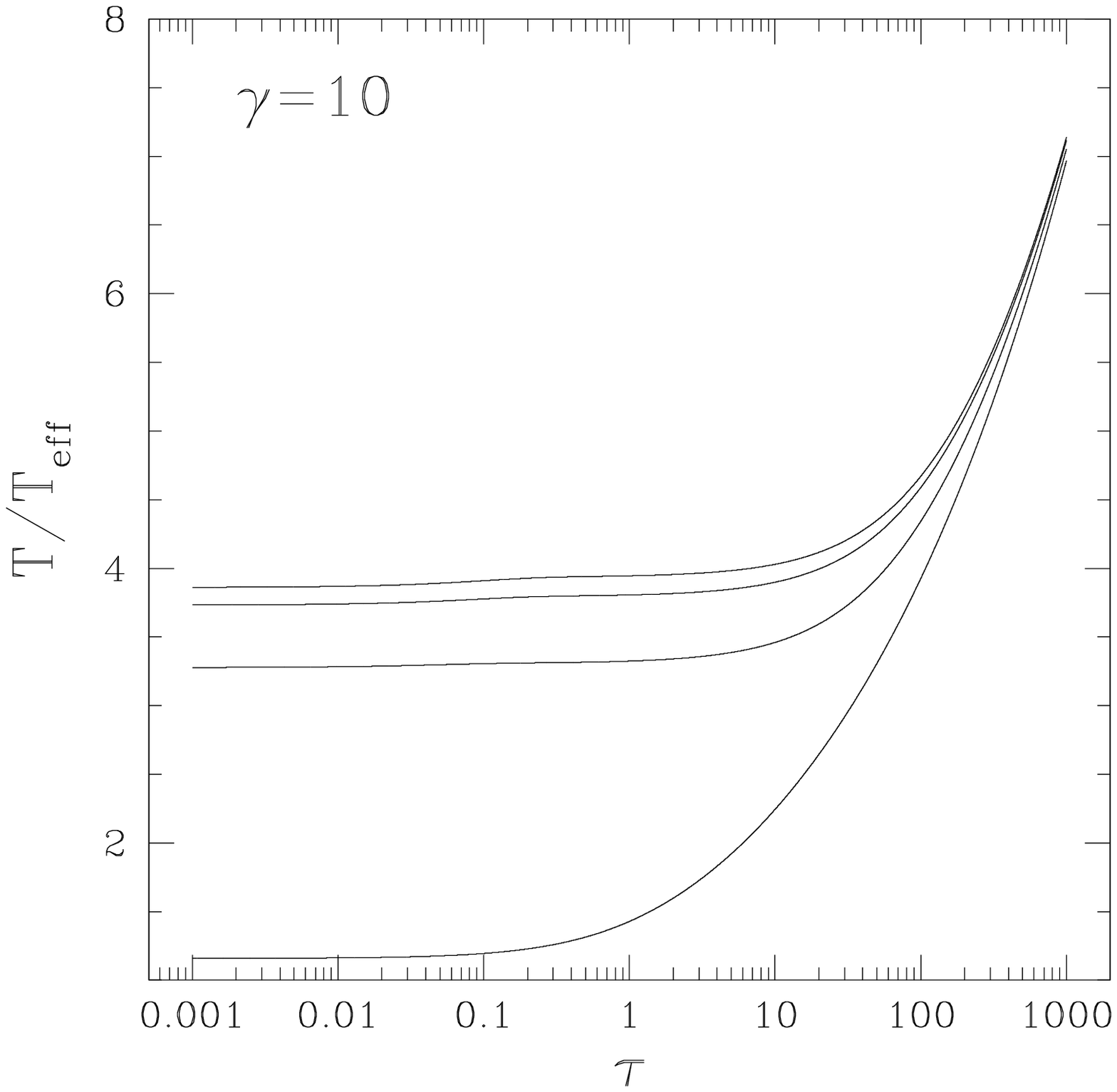}
\figcaption[f7.ps]{ The temperature profiles for angles of 0,30,60 and 90$^{\circ}$ to
the line of sight, for the case $\gamma=10$, i.e. where the irradiation is absorbed high
in the atmosphere.
 We have also used $(T_{0}/T_{eff})^4 = 100$. \label{Tplot2}}
\clearpage
\plotone{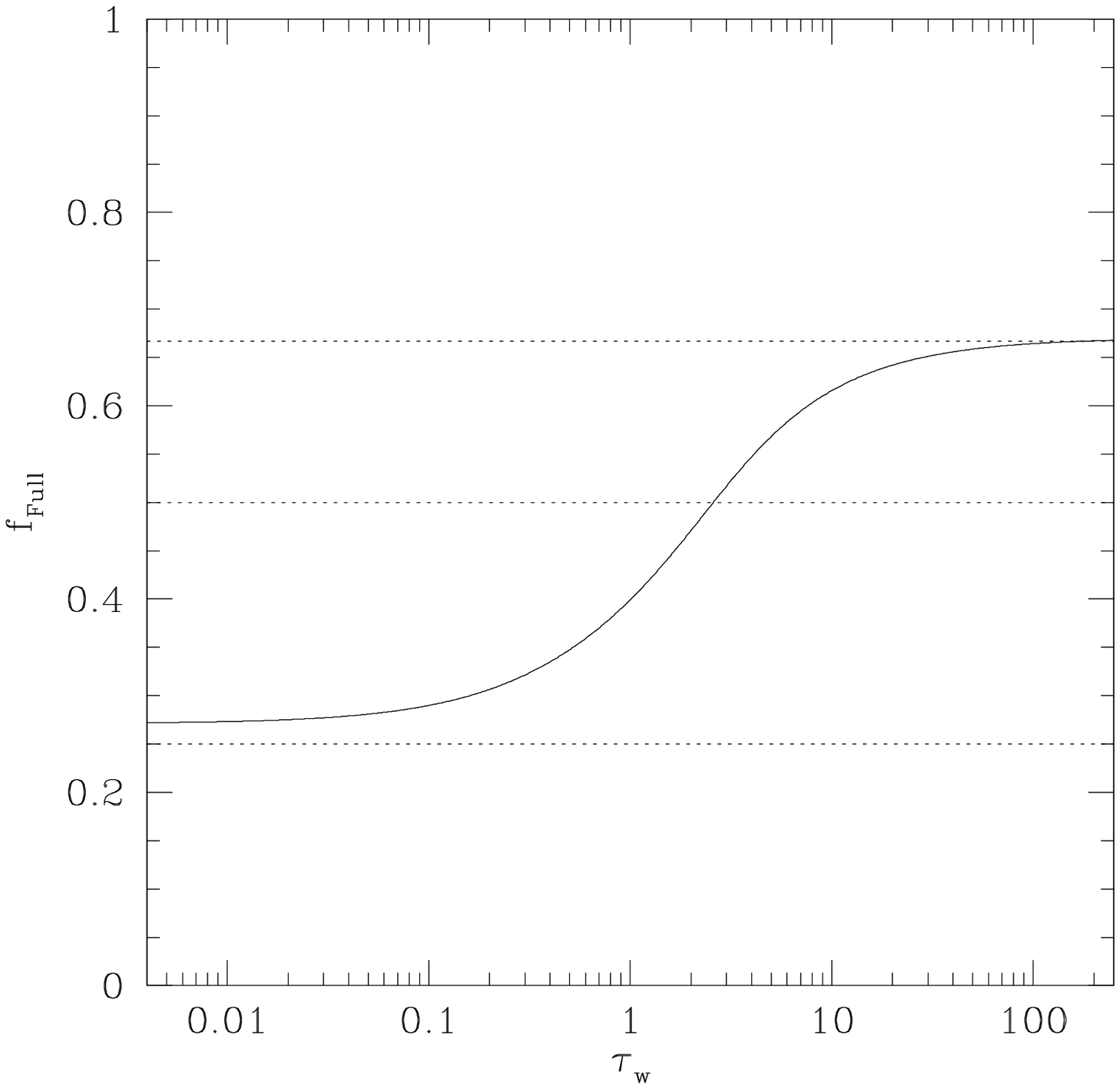}
\figcaption[f8.ps]{ The solid curve shows the effective $f$ value for the full phase flux
in our complete redistribution model, as a function of the redistribution depth $\tau_w$.
The dotted lines indicate the canonical values of 0.25, 0.5 and 2/3 discussed in the text. \label{Fullf}}
\clearpage
\plotone{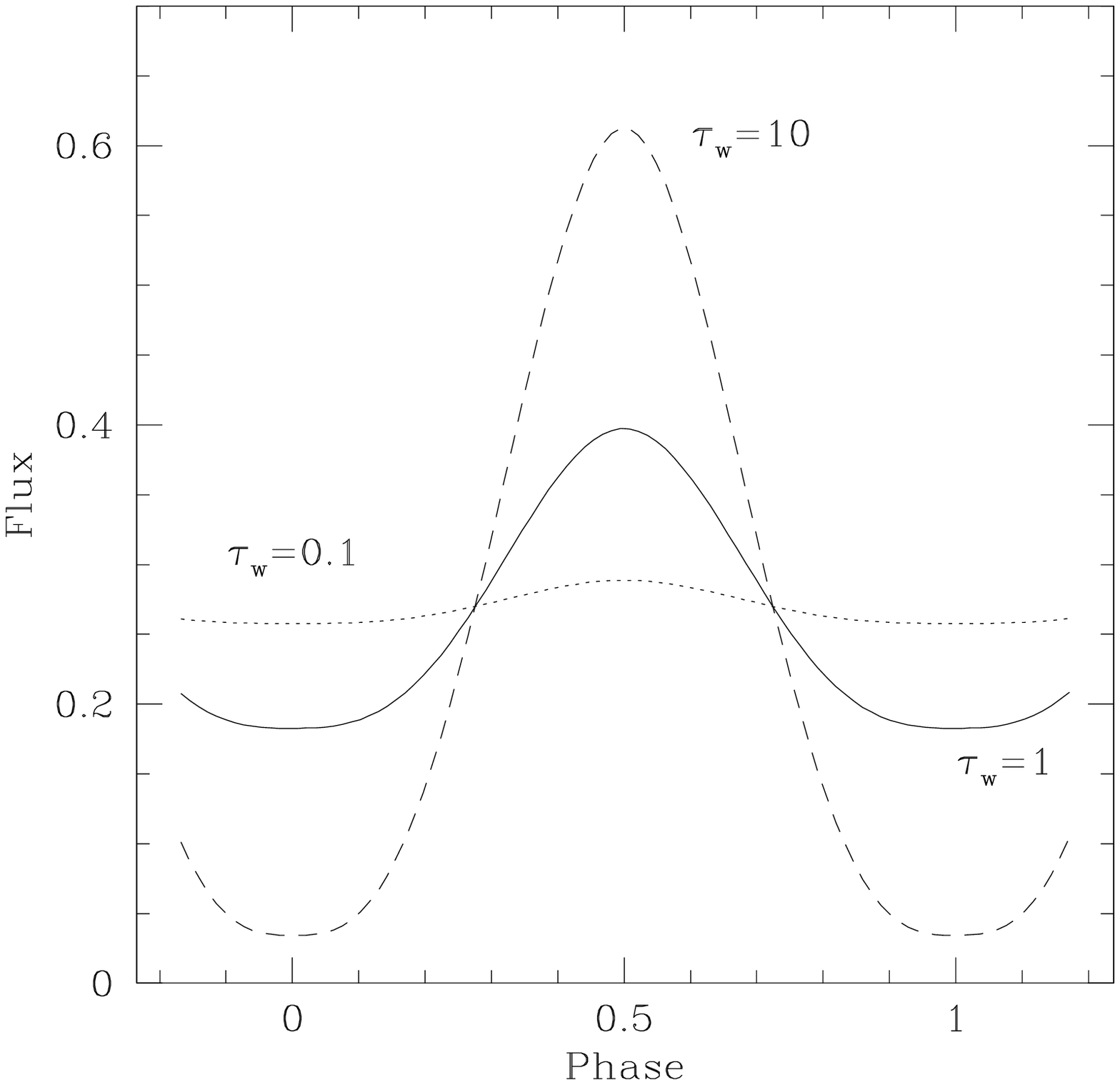}
\figcaption[f9.ps]{The solid line represents the phase curve for a planet in an edge-on orbit in which
the redistribution occurs at $\tau_w=1$. The dotted and dashed lines indicate the same, but for
$\tau_w=0.1$ and $\tau_w=10$. One can observe the transition from almost no redistribution to
almost complete uniformity as $\tau_w$ decreases. The flux level is parameterised in terms
of effective $f$. \label{RedPhase}}
\clearpage
\plotone{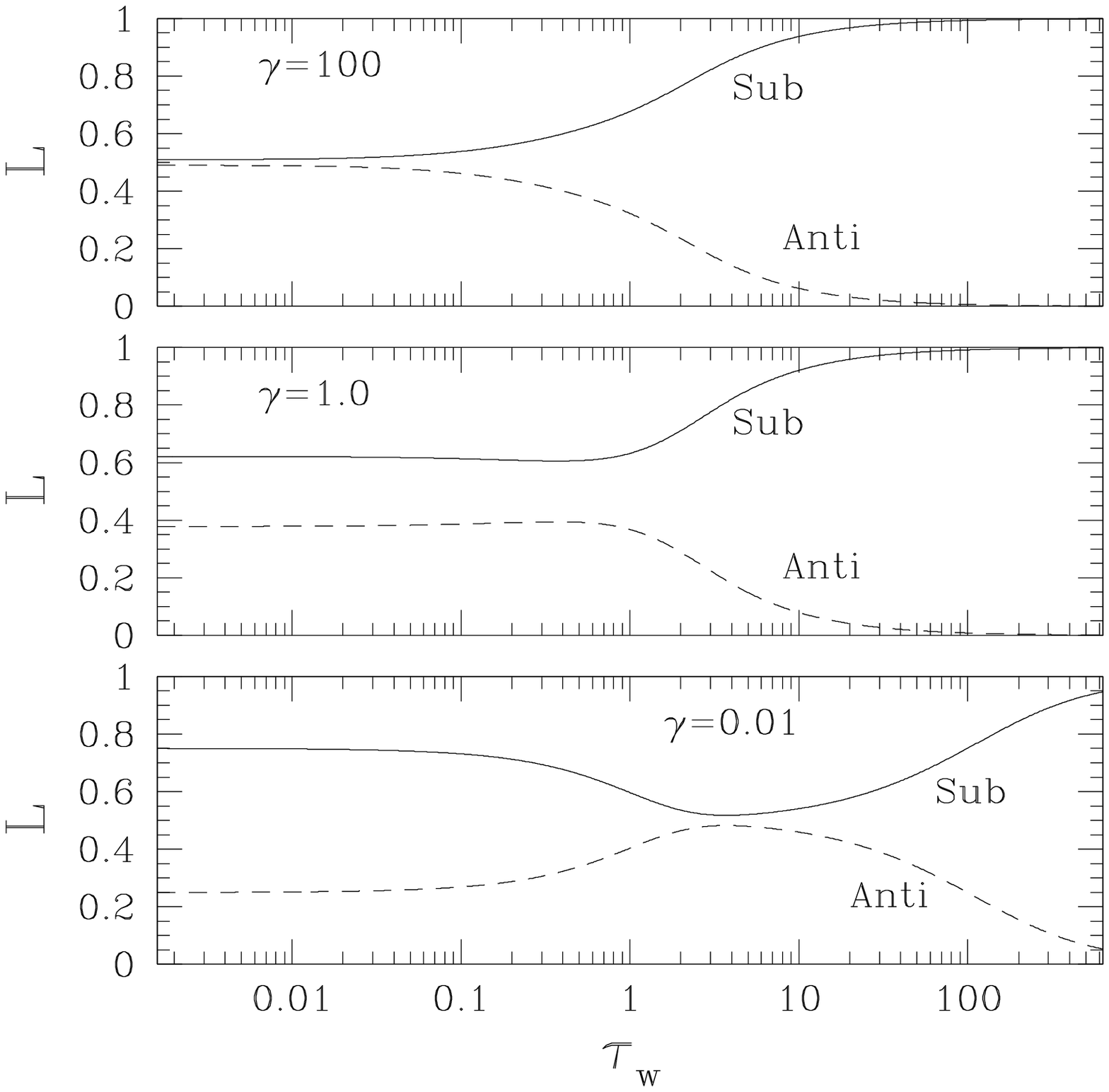}
\figcaption[f10.ps]{The three panels show how the relative luminosity of the substellar and antistellar
sides vary with $\tau_w$, for three different values of $\gamma$. For large $\gamma$ (upper panel), we
see a smooth transition from complete redistribution at low $\tau_w$ to essentially no redistribution
at large $\tau_w$. For lower values of $\gamma$ however, the dependance is more complicated.
\label{3Lg}}
\clearpage
\plotone{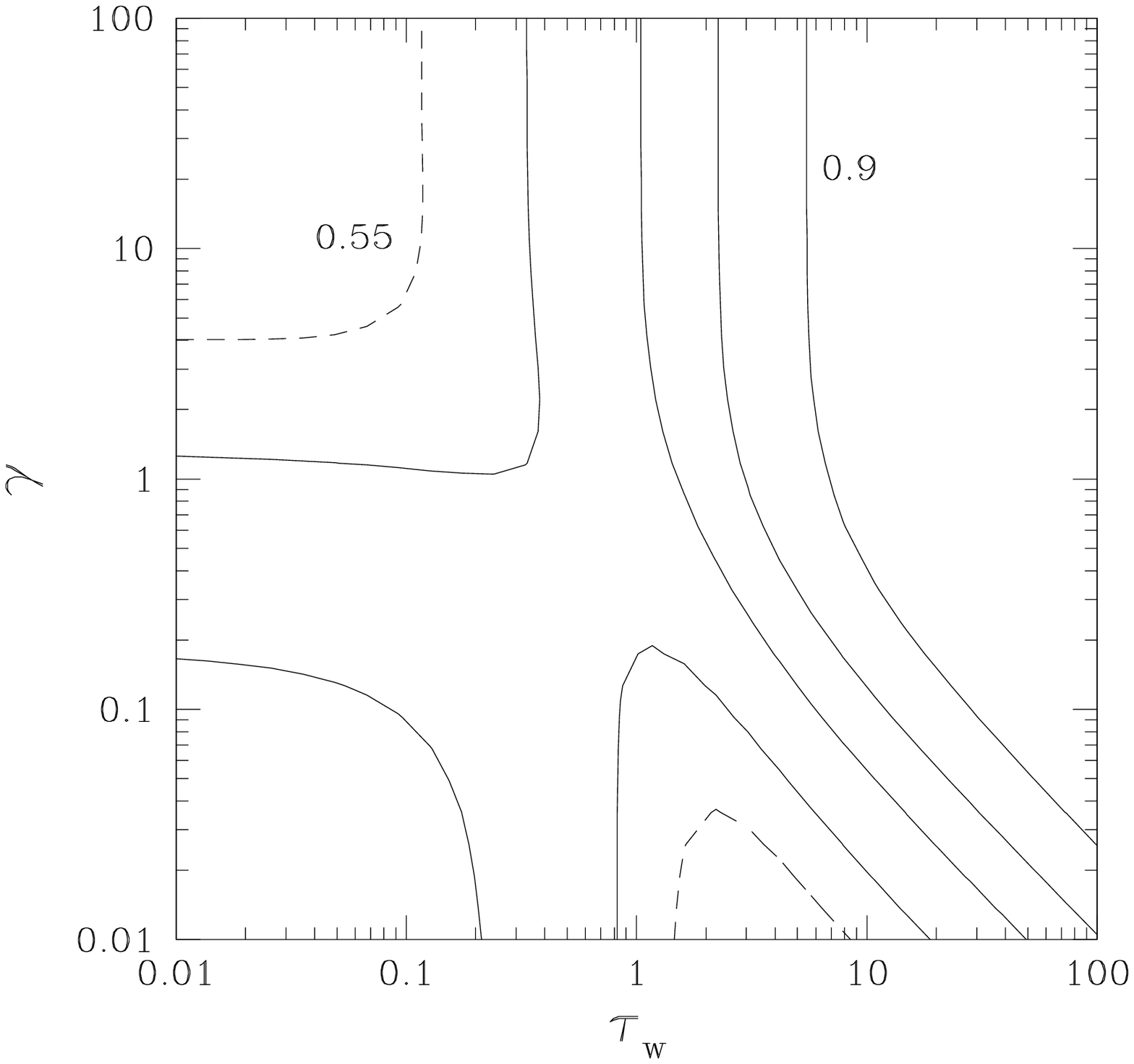}
\figcaption[f11.ps]{ The solid lines are contours of $L_{sub}/L_{total}$, ranging from 0.6 to 0.9.
The dashed contour is a value of 0.55. These latter contours enclose that region of the phase space where
the energy is approximately evenly redistributed.
\label{Lgt}}
\clearpage
\plotone{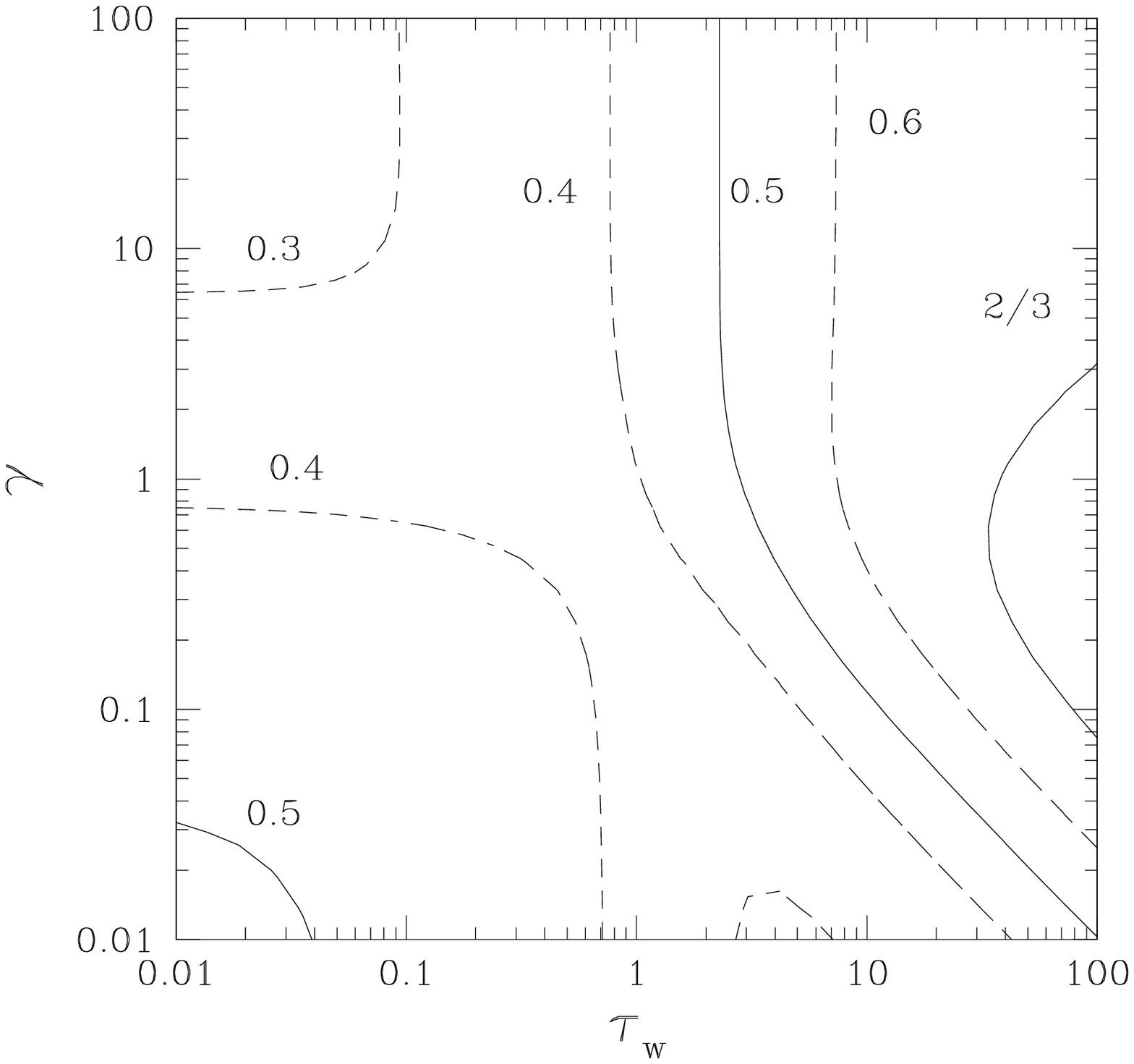}
\figcaption[f12.ps]{ The solid contours indicate values of the full-phase scaling parameter $f$ that
correspond to canonical values $1/2$ and $2/3$. Dashed curves indicate other values as labelled.
\label{fgt}}
\clearpage
\plotone{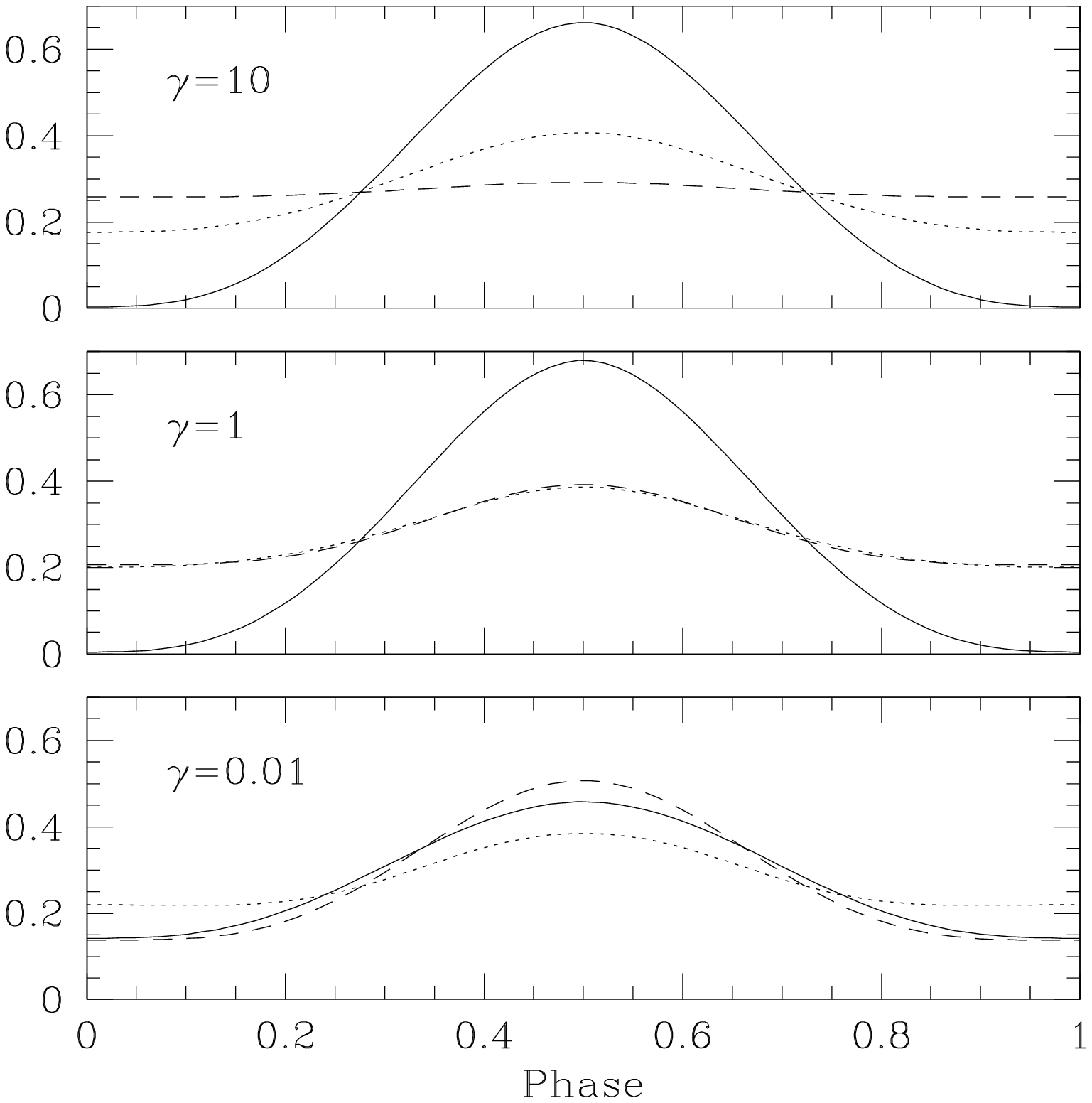}
\figcaption[f13.ps]{In each panel, corresponding to a different value of $\gamma$ as labelled,
the solid curve indicates $\tau_w=100$, the dotted curve $\tau_w=1$ and the dashed curve $\tau_w=0.01$.
In the middle panel, the dotted and dashed curves lie very close to one another. The nature of the
phase curve can be seen to depend clearly on both $\gamma$ and $\tau_w$.
\label{Phases}}
\clearpage
\plotone{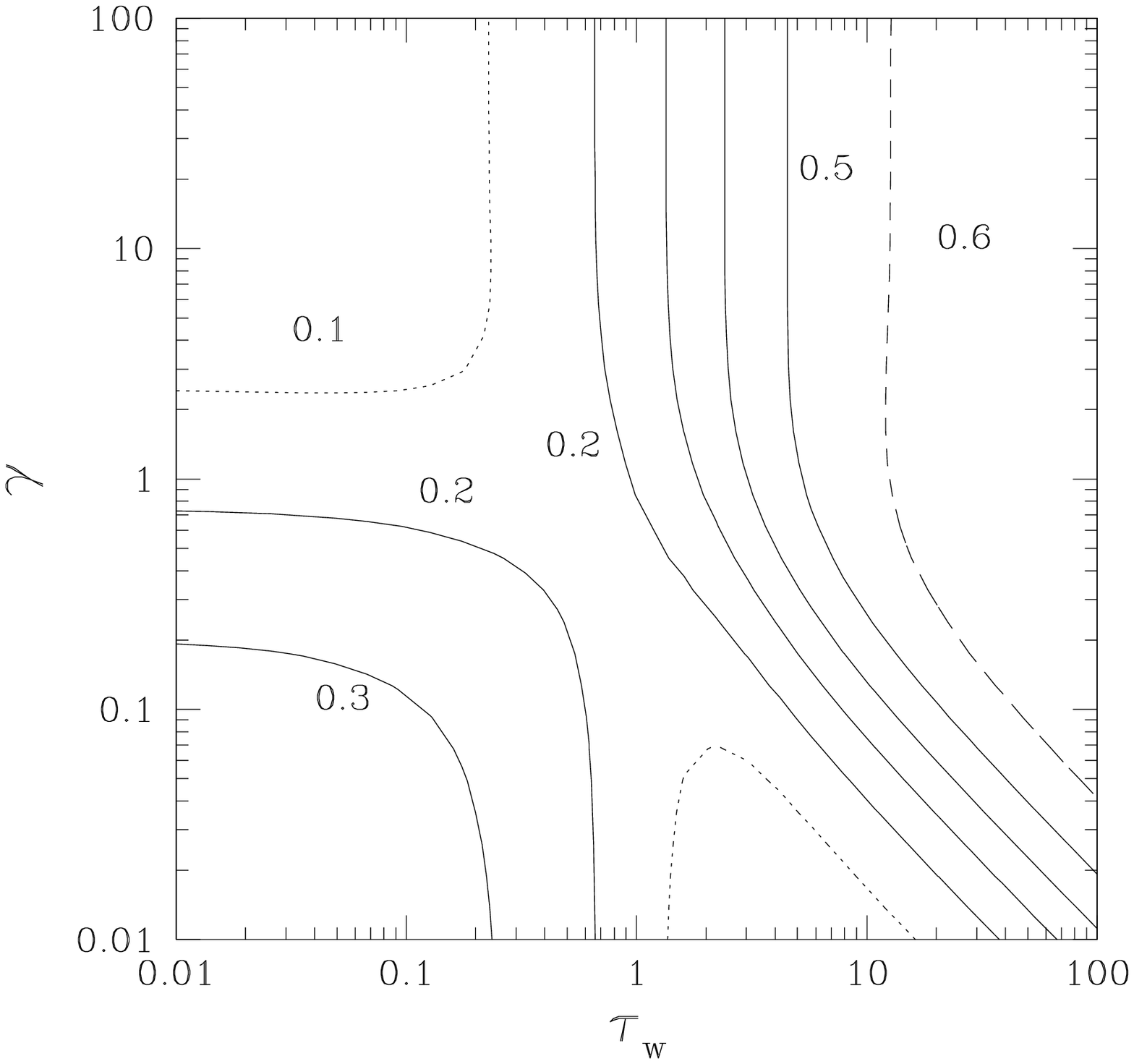}
\figcaption[f14.ps]{The contours plotted here are for the quantity $\delta f$, which is the difference
in $f_{eff}$ when observed at full phase and zero phase. We have assumed an edge-on orbit here.
The dotted and dashed curves represent $\delta f=0.6$ and $\delta f=0.1$ respectively.
The solid contours lie between these two values, in increments of 0.1. We see that phase variations of moderateamplitude are expected over a wide range of parameters.
\label{fphase}}
\clearpage
\plotone{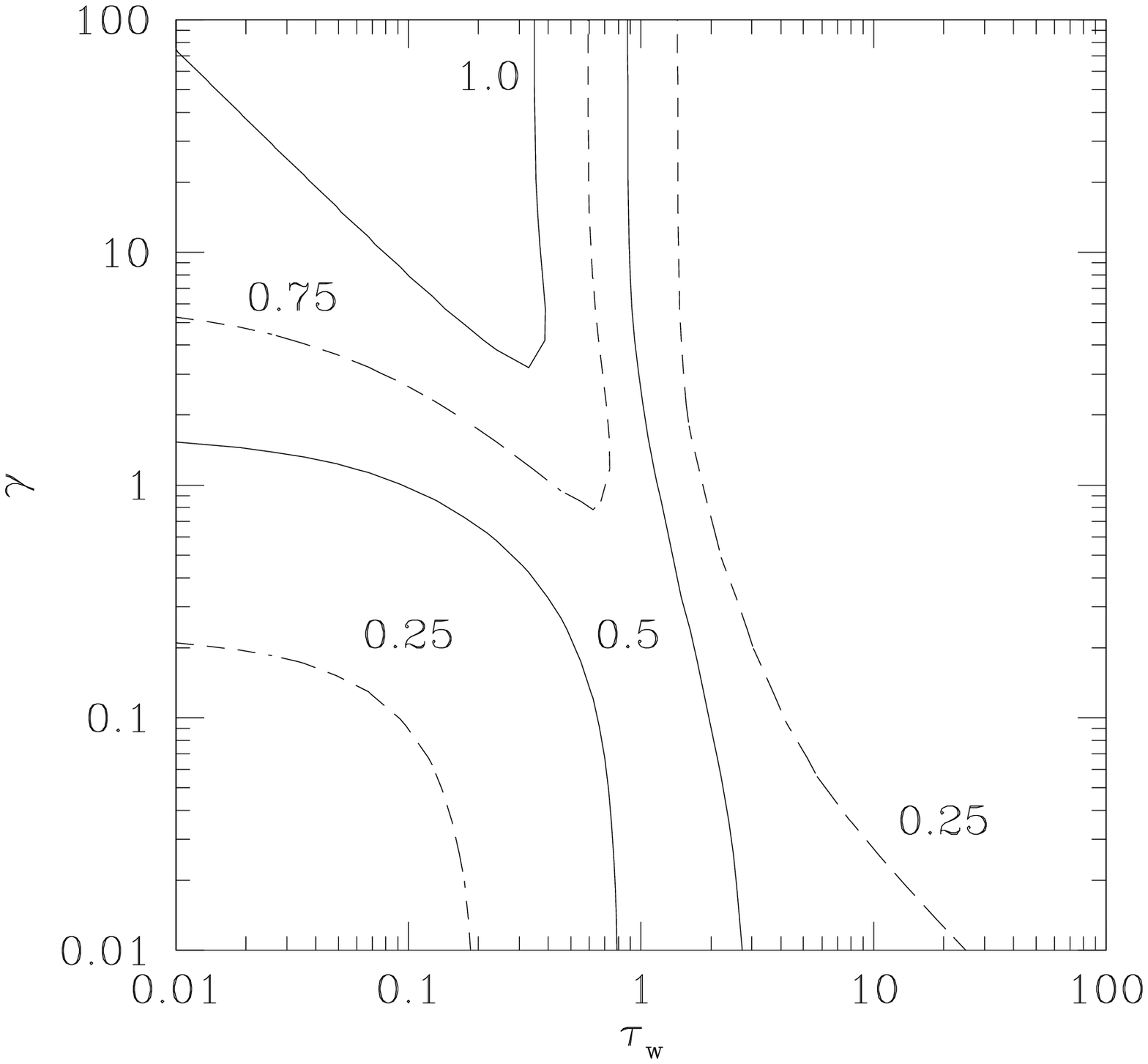}
\figcaption[f15.ps]{The contours indicate ratios of the intensity at the equator limb to the
intensity at the substellar point, when the planet is observed at full phase. The solid contours
indicate values of 0.5 and 1.0, while the dashed contours are 0.25 and 0.75. We see that deep
redistribution indicates strong limb darkening, while high altitude absorption with moderately
shallow redistribution can actually lead to limb brightening.
\label{RedLimb}}
\clearpage
\plotone{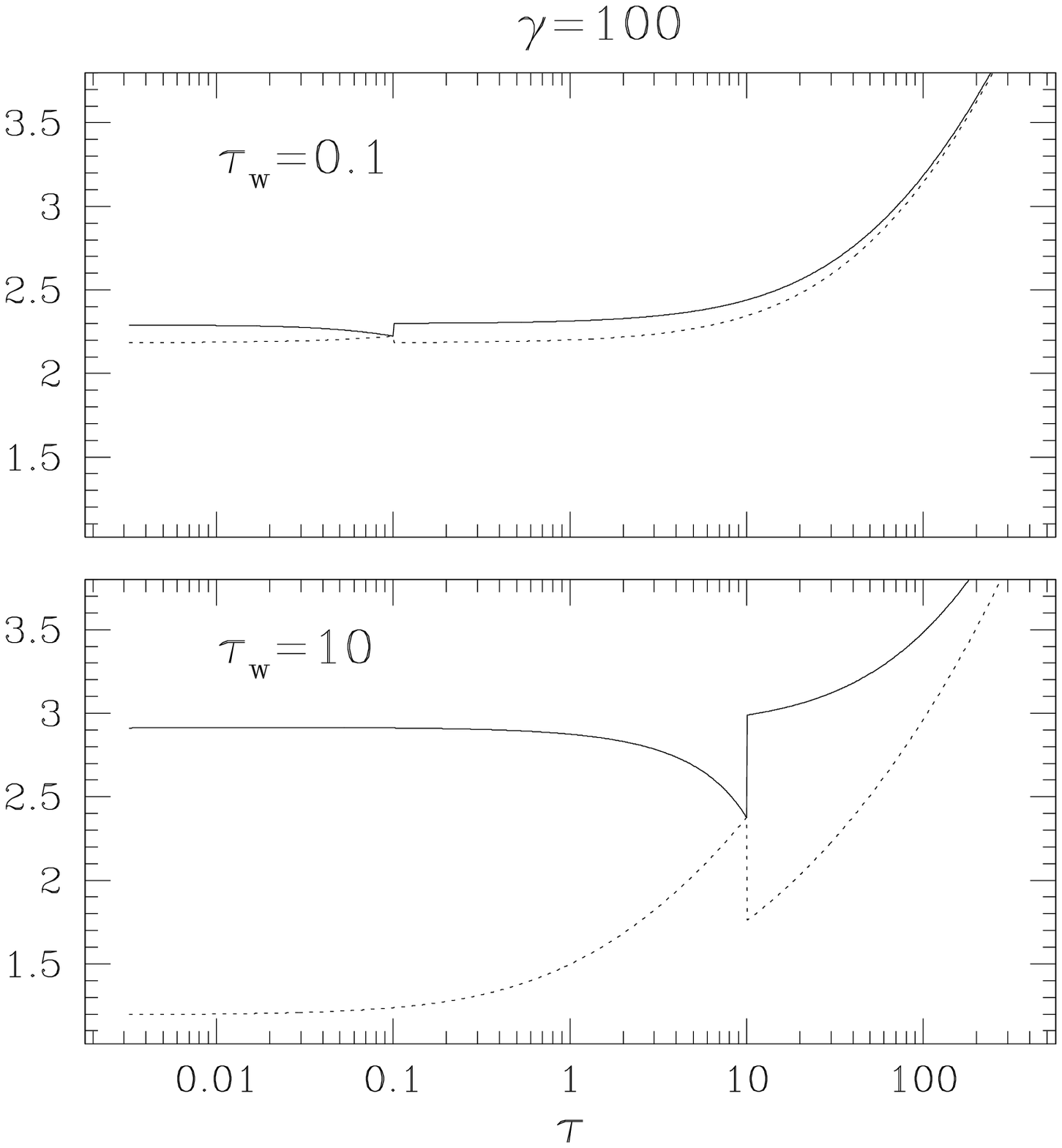}
\figcaption[f16.ps]{The solid lines show the temperature profile for the substellar point
in each case. The dotted lines indicate the temperature profile at the limb (for $\theta_p=\pi/2$).
The upper panel indicates redistribution is efficient in the case for low $\tau$ and large $\gamma$
but inefficient for large $\tau$ and large $\gamma$.
\label{Tprof1}}
\clearpage
\plotone{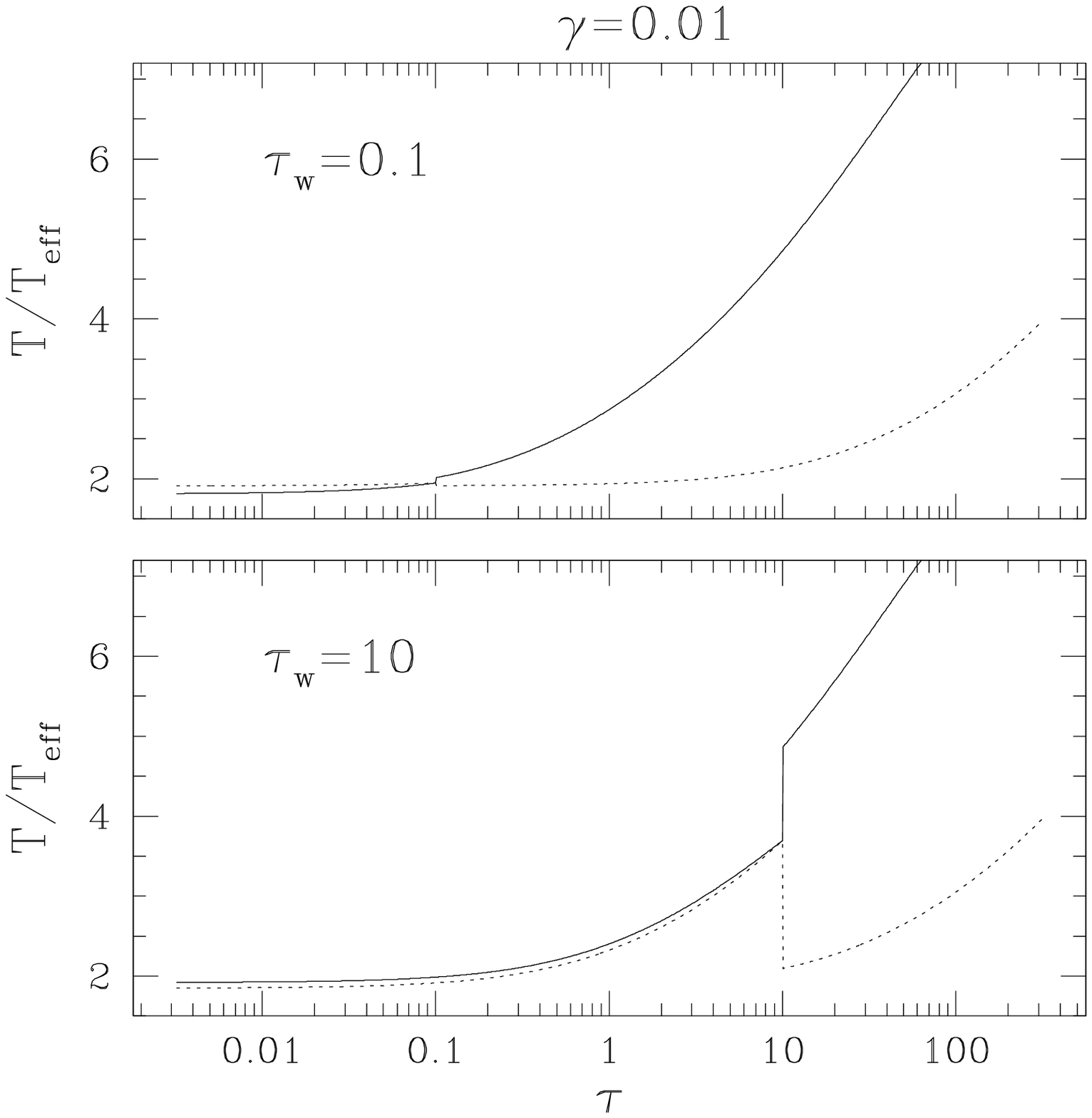}
\figcaption[f17.ps]{The solid lines show the temperature profile for the substellar point
in each case. The dotted lines indicate the temperature profile at the limb (for $\theta_p=\pi/2$).
The upper panel indicates redistribution is inefficient in the case for low $\tau$ and low $\gamma$
but efficient for large $\tau$ and small $\gamma$.
\label{Tprof2}}
\clearpage
\plotone{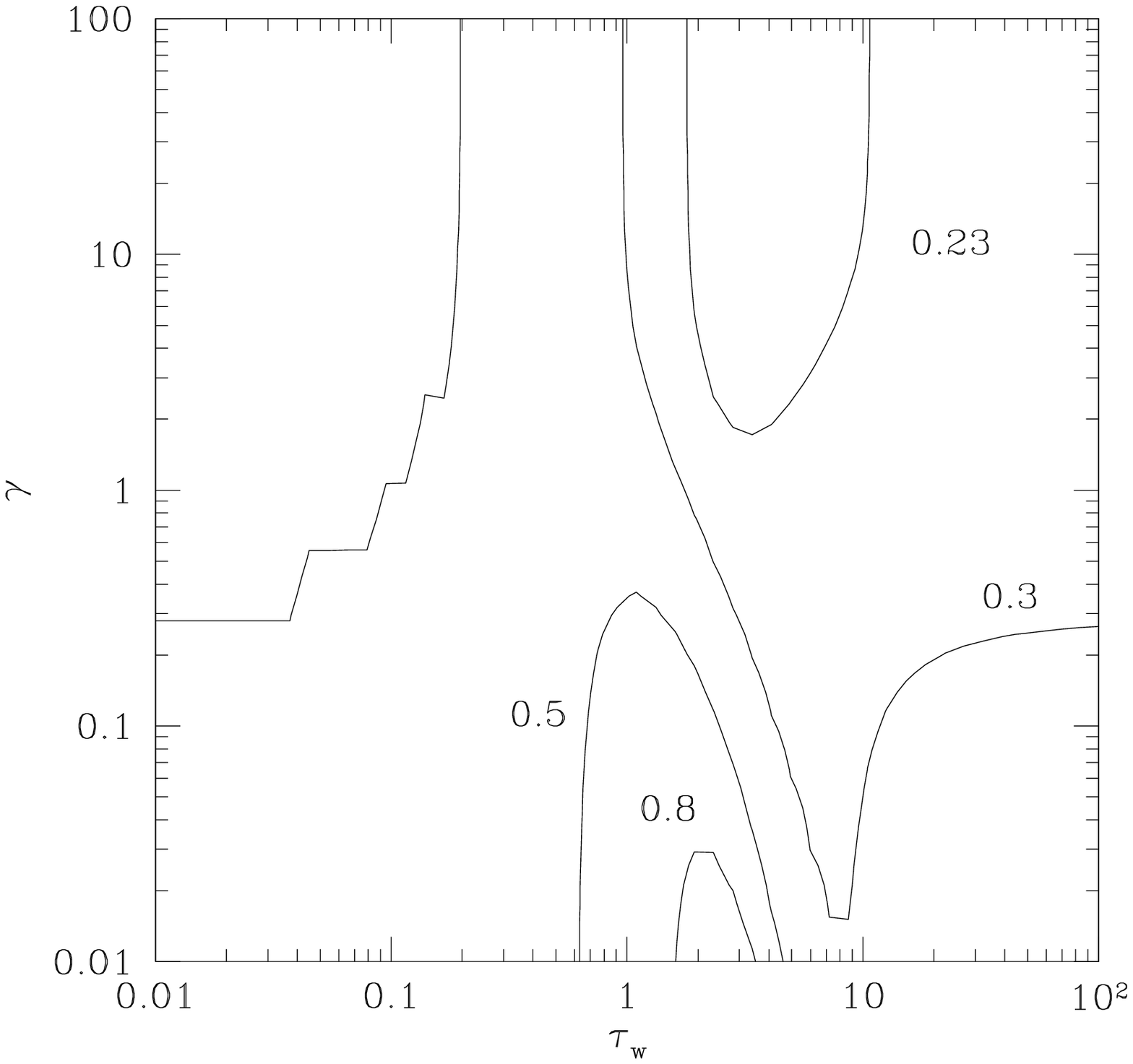}
\figcaption[f18.ps]{The solid contours indicate values of spectral line strength $A$. We see that
the strongest lines occur for $\gamma \ll 1$ and $\tau_w \sim 2$. For $\gamma >1$, the lines are
relatively weak and show little dependance on $\tau_w$.
\label{Line1}}
\clearpage
\plotone{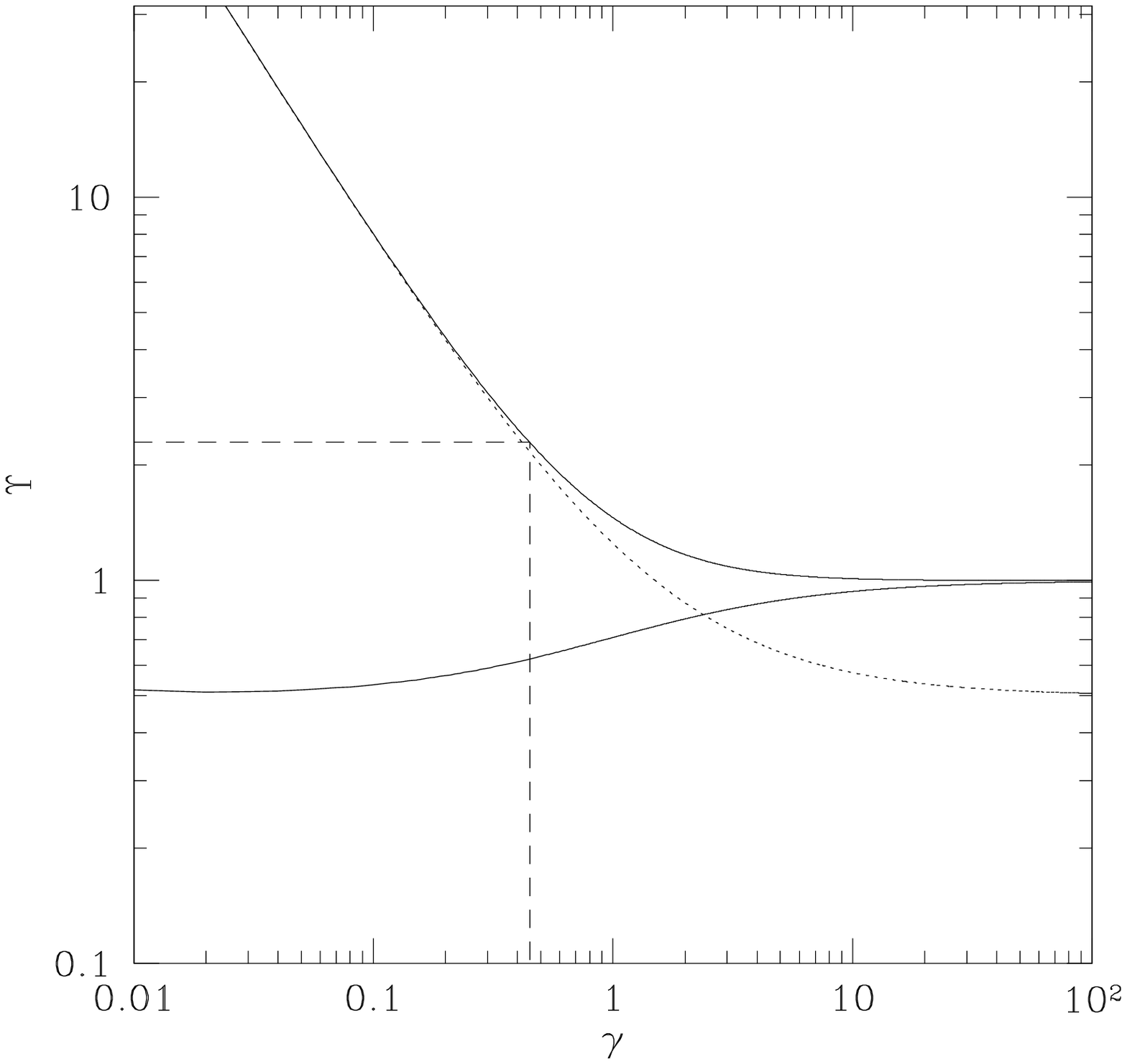}
\figcaption[f19.ps]{The upper solid line is the function $\Upsilon$($\gamma$). For $\gamma \ll 1$, the
function asymptotes to $\rightarrow 1 + 3/2\gamma$, indicated by the dotted line. The dashed line corresponds to the solution for
the Barman et al. (2005) case. The lower solid line is the corresponding function for the lower temperature plateau
in our simple model.
\label{Func}}
\clearpage
\plotone{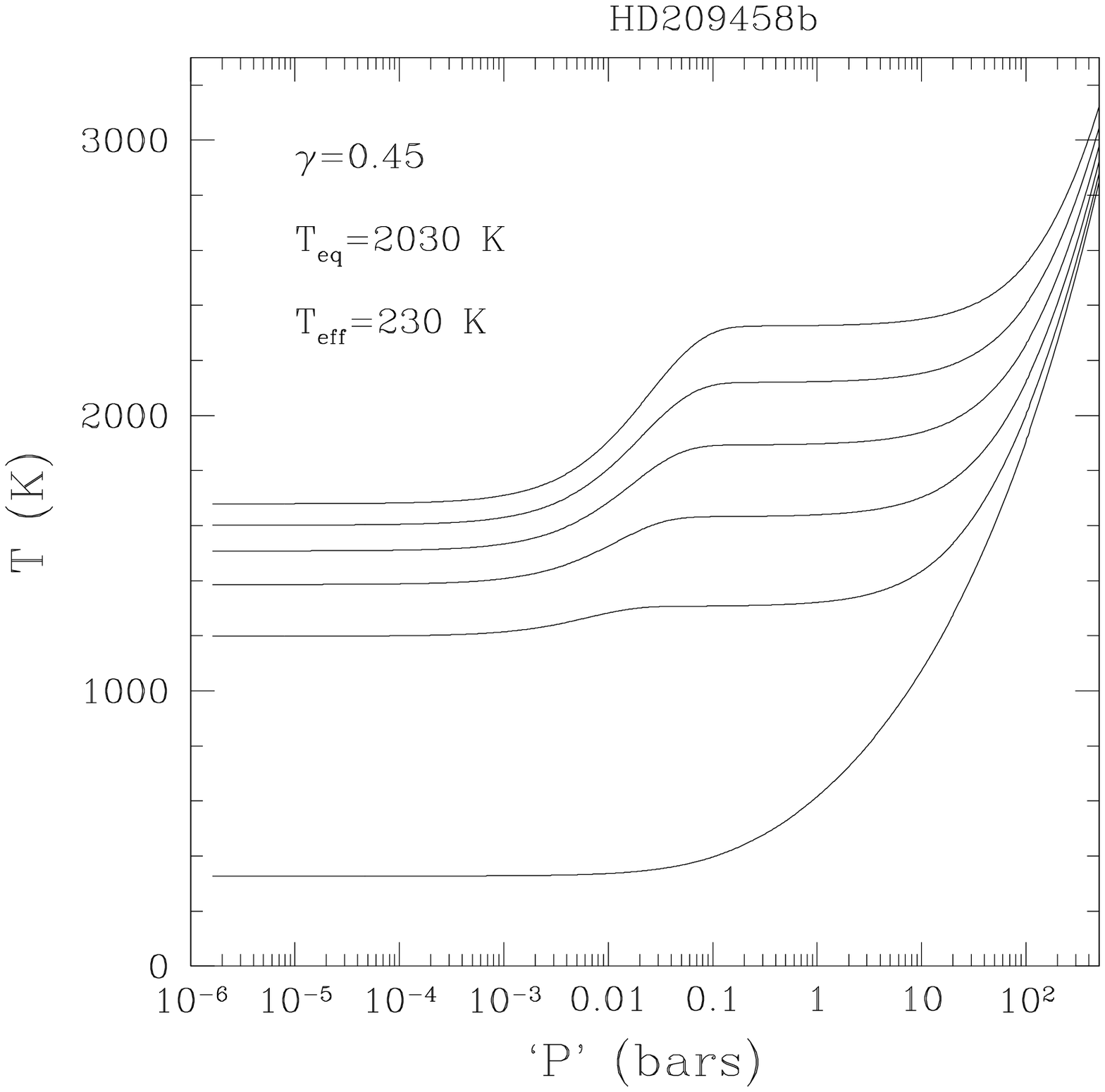}
\figcaption[f20.ps]{ The solid curves show the temperature profiles for $\mu_0 = 1.0$,0.8,0.6,0.4,0.2 and 0
(from top to bottom). The value of $\gamma=0.45$ was chosen to match the normalisation of the models of
Barman et al. (2005). The other parameters $T_{0}$ and $T_{eff}$ were chosen based on the values used
by Barman et al. The
horizontal axis is really $\tau$, but we have scaled this down by a factor 63 to mimic pressure in bars.
(We expect the scaling to be linear for a plane parallel atmosphere with constant opacity. In fact, using
the known gravity of HD~209458b, we can even derive the effective mean IR opacity $\sim 0.07 cm^2 g^{-1}$.
\label{BHA}}
\clearpage
\clearpage
\plotone{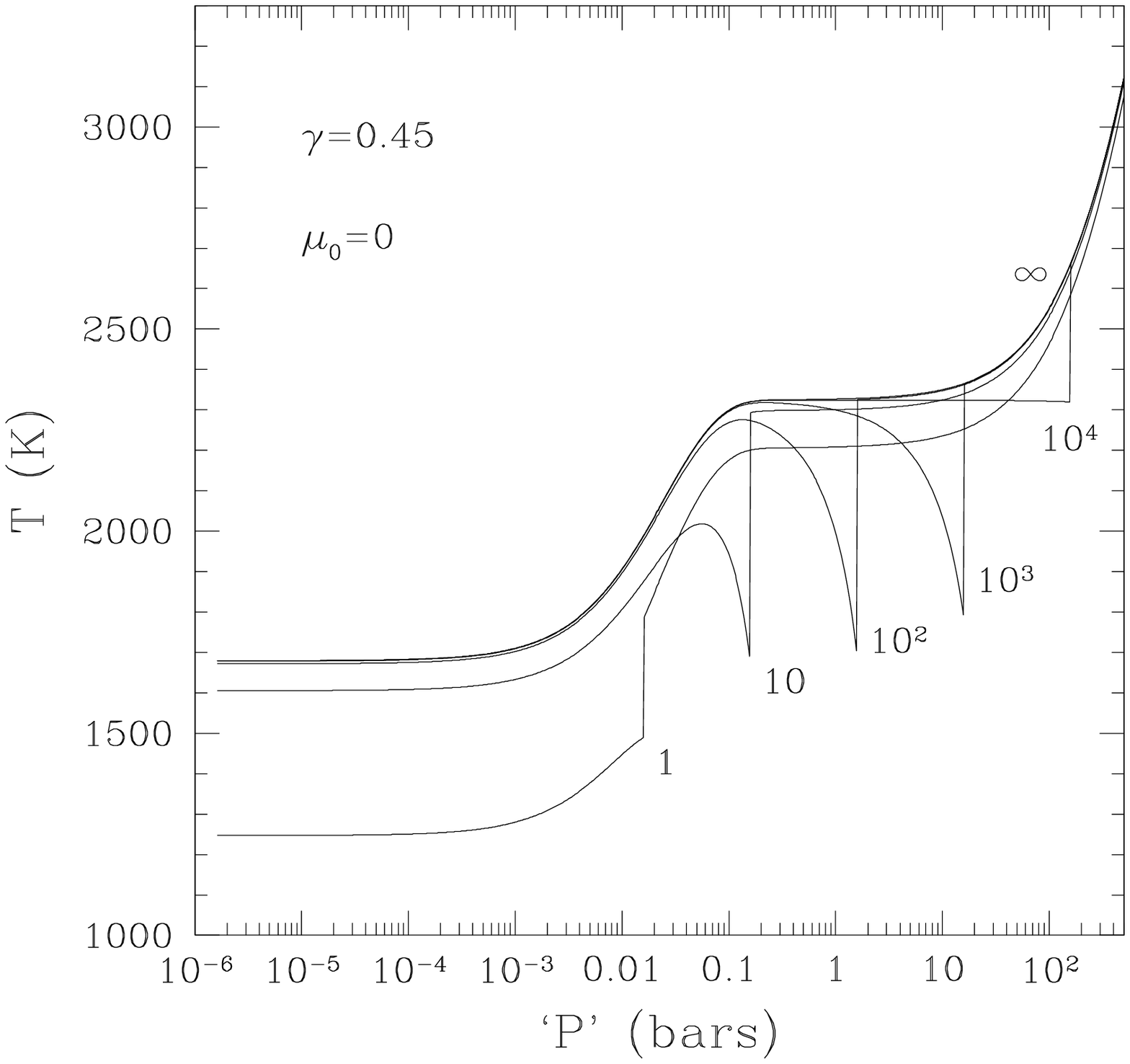}
\figcaption[f23.ps]{ The curves show how the temperature profile of the $\gamma=0.45$ model at the
substellar point changes as we decrease $\tau_w$, the optical depth of the redistribution layer.
Each curve is labelled with the appropriate value of $\tau_w$. The effect of redistribution at the
hottest point on the star is clearly to remove energy and transport it to cooler parts of the star.
Thus, we see depressions in the temperature profile near $\tau_w$. However, for values of $\tau_w \sim 10^2$--$10^4$, we see that this region lies in the middle of the temperature plateau where much of the energy is
being absorbed. The combination of these two effects leads to conspicuous temperature bumps.
\label{BH3}}
\clearpage
\plotone{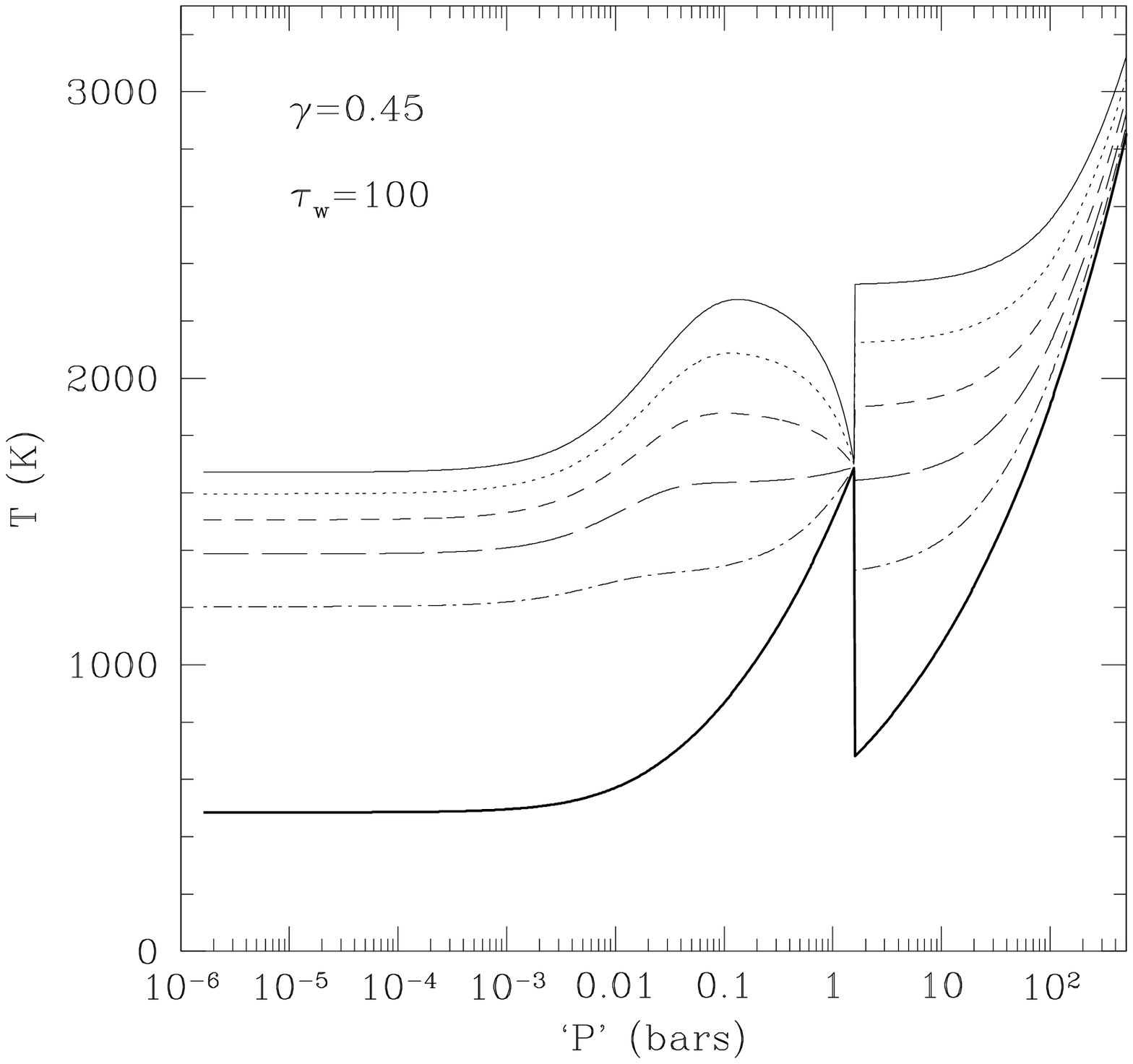}
\figcaption[f24.ps]{ The curves show how the temperature profile of the $\gamma=0.45$ model varies
along the equator from center to limb, assuming $\tau_w = 100$. The curves are for values of
$\mu_0$ running from 1.0 (top) to 0.0 (bottom), in increments of 0.2. The heavy solid line is
the profile at the limb, which also applies on the nonirradiated antistellar side of the planet.
We see that the removal of energy by irradiation near the center changes into an injection of
energy near the limb and on the antistellar side.
\label{BH6}}
\clearpage
\plotone{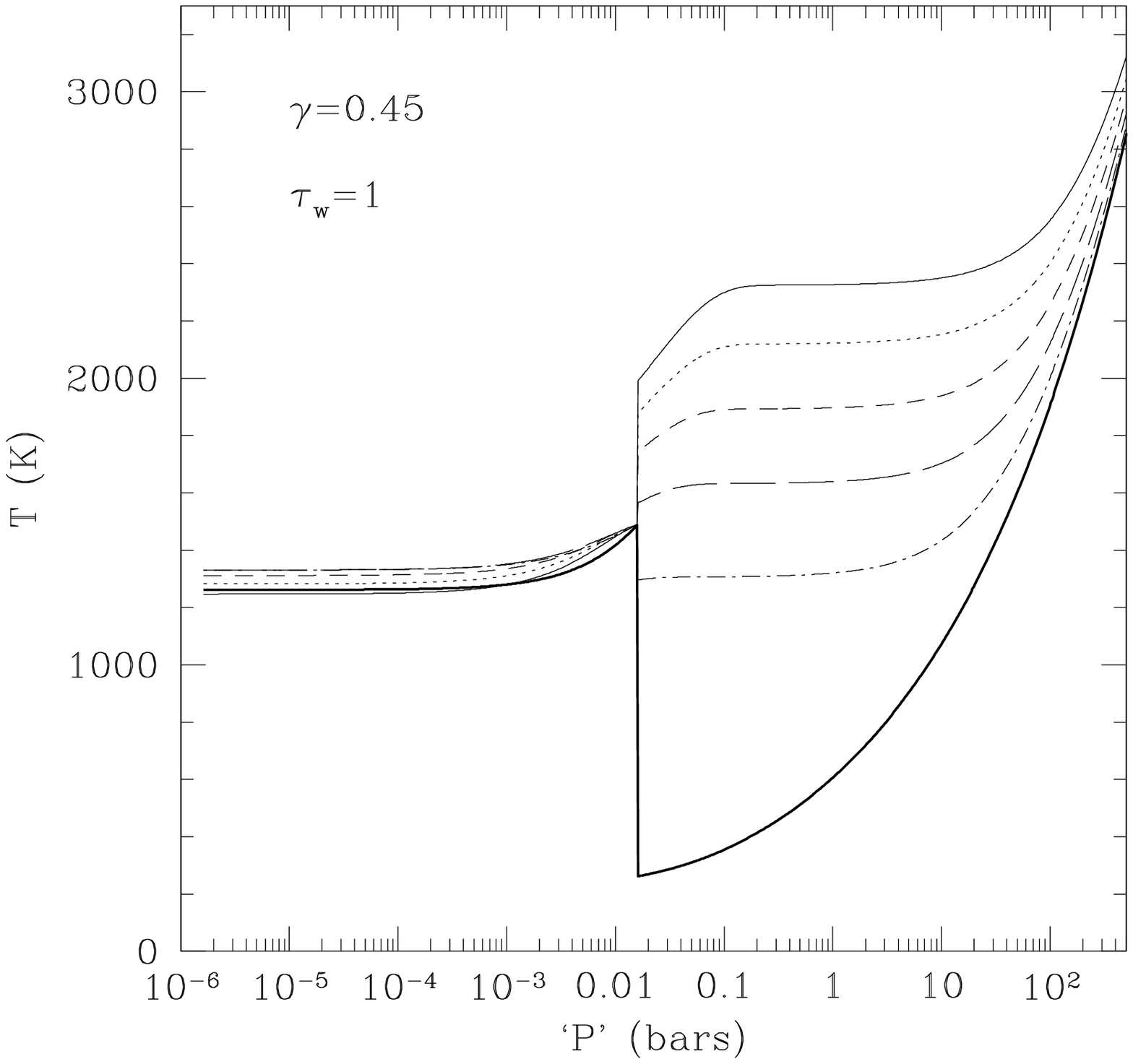}
\figcaption[f24a.ps]{ The curves show how the temperature profile of the $\gamma=0.45$ model varies
along the equator from center to limb, assuming $\tau_w = 1$. The curves are for values of
$\mu_0$ running from 1.0 (top) to 0.0 (bottom), in increments of 0.2. The heavy solid line is
the profile at the limb, which also applies on the nonirradiated antistellar side of the planet.
We see that the removal of energy by irradiation near the center changes into an injection of
energy near the limb and on the antistellar side, with a net result of enforcing a very uniform
temperature at lower pressures.
\label{BH7}}
\clearpage
\plotone{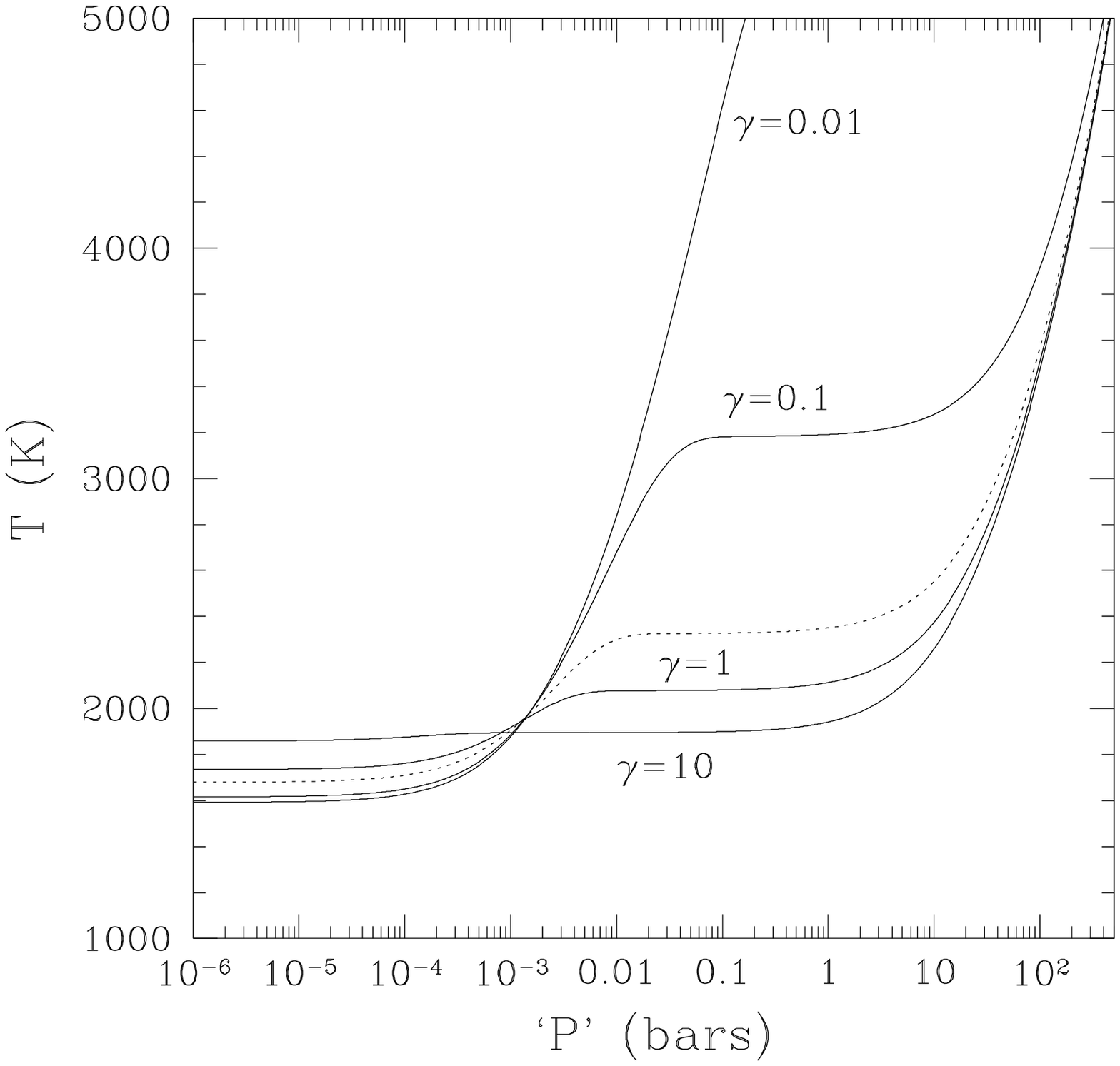}
\figcaption[f25.ps]{ The curves show the temperature profiles at the substellar point ($\mu_0=1$) as
we increase $\gamma$ from a value of 0.01 to $\gamma=10$. As $\gamma$ increases, the
incoming energy is absorbed at lower and lower pressures, and the temperature jump decreases, eventually
leading to an almost isothermal profile. The dotted line is for the value $\gamma=0.45$, as calibrated
for Figure~\ref{BHA}.
\label{BH2}} 
\clearpage
\plotone{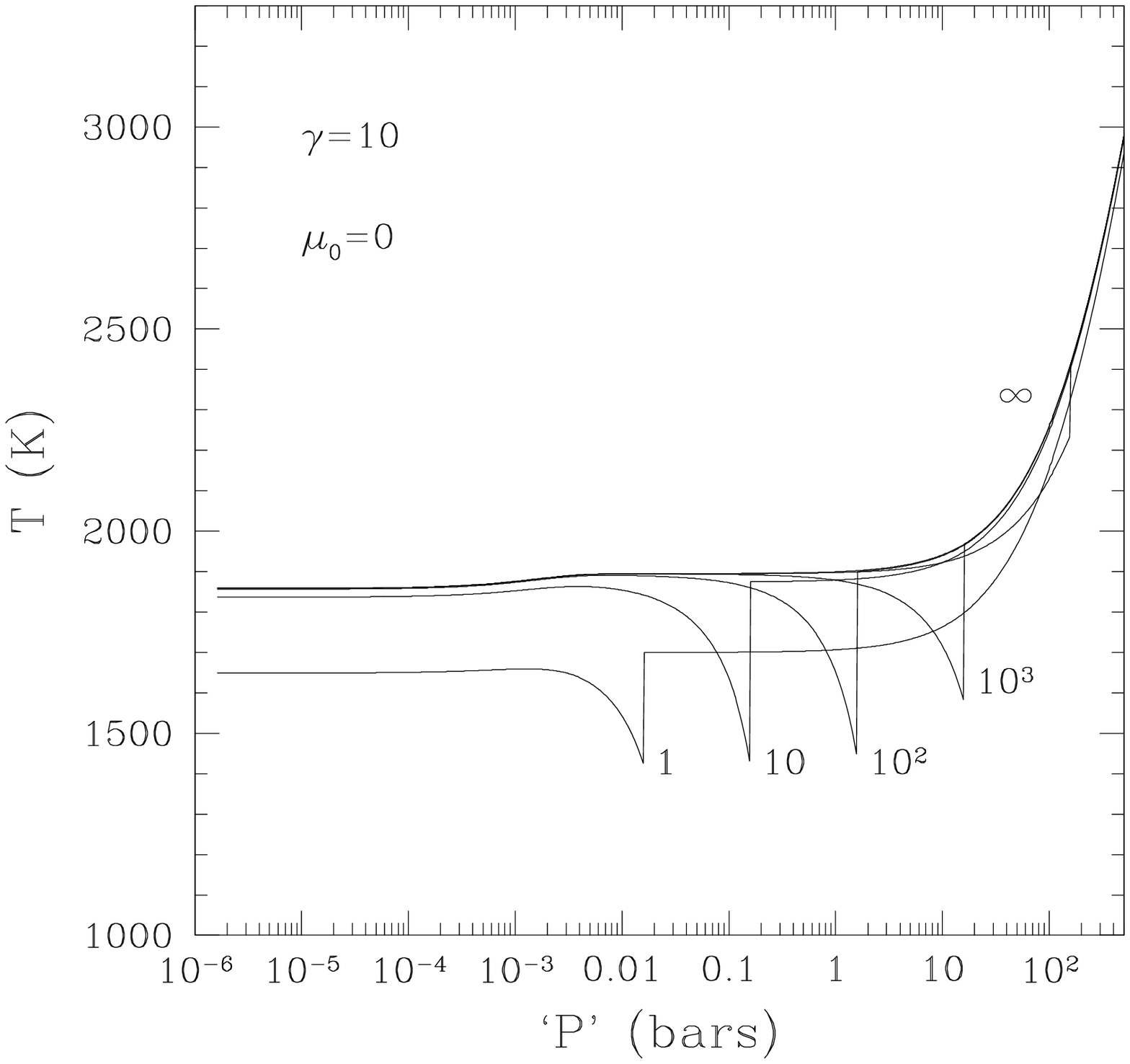}
\figcaption[f26.ps]{Redistribution in the $\gamma=10$ model is not quite as dramatic, although
we still find temperature inversions, which get deeper as $\tau_w$ decreases. Recall that $\gamma \sim 10$
and $\tau_w \sim 3$ is a good model for the HD209458b case, suggesting that this atmosphere may have
a temperature inversion at the 20~millibar level.
\label{BH4}}

\begin{references}
\reference{BSH} Ballester, Sing \& Herbert, 2007, Nature, 445, 511
\reference{BHA0} Barman, T. S., Hauschildt, P. H. \& Allard, F., 2001, ApJ, 556, 885
\reference{BHA} Barman, T. S., Hauschildt, P. H. \& Allard, F., 2005, ApJ, 632, 1132
\reference{B07} Barman, T. S., 2007, ApJ, 661, L191
\reference{BCB} Baraffe, I., Chabrier, G., Barman, T., Allard, F. \& Hauschildt, P., 2003, A\&A, 402, 701
\reference{Bouch} Bouchy, F., et al., 2005, A\&A, 444, L15
\reference{B01} Brown, T. M., Charbonneau, D., Gilliland, R. L., Noyes, R. \& Burrows, A., 2001, ApJ, 552, 699
\reference{BSH} Burrows, A., Sudarsky, D. \& Hubbard, W. B., 2003, ApJ, 594, 545
\reference{BHS} Burrows, A., Hubeny, I. \& Sudarsky, D., 2005, ApJ, 625, L135
\reference{BHB} Burrows, A., Hubeny, I., Budaj, J., Knutson, H. A., \& Charbonneau, D., 2007. ApJ, 668, L171
\reference{BSH} Burrows, A., Sudarsky, D. \& Hubeny, I., 2006, ApJ, 650, 1140
\reference{C00} Charbonneau, D., Brown, T., Latham, D. \& Mayor, M., 2000, ApJ, 529, L45
\reference{C02} Charbonneau, D., Brown, T. M., Noyes, R. W. \& Gilliland, R., 2002, ApJ, 568, 377
\reference{C05} Charbonneau, D., et al., 2005, ApJ, 626, 523
\reference{Ch} Chauvin, G., Lagrange, A., Dumas, C., Zuckerman, B., Mouillet, D., Song, I., Beuzit, J.-L. \& Lowrance, P., 2005, A\&A, 438, L25
\reference{Cho1} Cho, J., Menou, K., Hansen, B. \& Seager, S., 2003, ApJ 587, L117
\reference{Cho2} Cho, J., Menou, K., Hansen, B. \& Seager, S., 2008, ApJ, in press; astro-ph/0607338
\reference{CS} Cooper, C. S. \& Showman, A. P., 2005, ApJ, 629, L45
\reference{CS2} Cooper, C. S. \& Showman, A. P., 2006, ApJ, 649, 1048
\reference{DSRH} Deming, D., Seager, S., Richardson, L. J. \& Harrington, J., 2005, Nature, 434, 740
\reference{DHSR} Deming, D., Harrington, J., Seager, S. \& Richardson, L. J., 2006, ApJ, 644, 560
\reference{Dyu} Dyudina, U. A., Sackett, P., Bayliss, D. R., Seager, S., Porco, C. C., Throop, H. B. \& Dones, L., 2005, ApJ, 618, 973
\reference{FML} Fortney, J. J., Marley, M. S., Lodders, K., Saumon, D. \& Freedman, R., 2005, ApJ, 627, L69
\reference{Fsolo} Fortney, J. J., 2005, MNRAS, 364, 649
\reference{FSM} Fortney, J. J., Saumon, D., Marley, M. S., Lodders, K., \& Freedman, R. S., 2006, ApJ, 642, 495
\reference{Fetc} Fortney, J. J., Cooper, C. S., Showman, A. P., Marley, M. S. \& Freedman, R. S., 2006, ApJ, 652, 746
\reference{F07} Fortney, J. J., Lodders, K., Marley, M. S. \& Freedman, R. S., 2007, arXiV:0710.2558
\reference{Grill} Grillmair, C. J., Charbonneau, D., Burrows, A., Armus, L., Stauffer, J., Meadows, V., Van Cleve, J. \& Levine, D., 2007, ApJ, 658, L115
\reference{Gui} Guillot, T., Burrows, A., Hubbard, W. B., Lunine, J. I. \& Saumon, D., 1996, ApJ, 459, L35
\reference{HH} Harrington, J., Hansen, B. M., Luszcz, S. H., Seager, S, Deming, D., Menou, K., Cho, J. Y.-K., \& Richardson, L. J., 2006, Science, 314, 623
\reference{HL} Harrington, J., Luszcz, S., Seager, S., Deming, D. \& Richardson, L. J. 2007, Nature, 447, 691
\reference{Hen} Henry, G. W., Marcy, G. W., Butler, R. P. \& Vogt, S. S., 2000, ApJ, 529, L41
\reference{HBS} Hubeny, I., Burrows, A. \& Sudarsky, D., 2003, ApJ, 594, 1011
\reference{Iro} Iro, N., B\'{e}zard, B. \& Guillot, T., 2005, A\&A, 436, 719
\reference{Knu} Knutson, H. A., Charbonneau, D., Noyes, R. W., Brown, T. M. \& Gilliland, R. L., 2007a, ApJ, 655, 564
\reference{Kb} Knutson, H. A., et al., 2007b, Nature, 447, 183
\reference{Kc} Knutson, H. A., Charbonneau, D., Allen, L. E., Burrows, A., \& Megeath, T. S., 2007c, arXiV:0709.3984
\reference{LL} Langton, J. \& Laughlin, G., 2007, ApJ, 657, L113
\reference{LR} Lucas, P. W., \& Roche, P. F., 2000, MNRAS, 314, 858
\reference{MFSB} Marley, M. S., Fortney, J., Seager, S. \& Barman, T., 2007, Protostars and Planets V, B. Reipurth, D. Jewitt, and K. Keil (eds.), University of Arizona Press, Tucson, 951, p733; 
astro-ph/0602468
\reference{Men} Menou, K., Cho J., Seager S., \& Hansen B., 2003, ApJ, 587, L113
\reference{M78} Mihalas, D., 1978, `Stellar Atmospheres', W. H. Freeman \& Co.
\reference{Mil} Milne, E. A., 1926, MNRAS, 87, 43
\reference{Neu} Neuhauser, R., Guenther, E. W., Wuchterl, G., Mugrauer, M., Bedalov, A. \& Hauschildt, P. H., 2005, A\&A, 435, L13
\reference{Raua} Rauscher, E., Menou, K., Seager, S., Deming, D., Cho, J. \& Hansen, B., 2007a, ApJ, 664, 1199
\reference{Raub} Rauscher, E., Menou, K., Seager, S.,  Cho, J. \& Hansen, B., 2007b, ApJ, 662, L115
\reference{R06} Richardson, L. J., Harrington, J., Seager, S., \& Deming, D., 2006, ApJ, 649, 1043
\reference{R07} Richardson, L. J., Deming, D., Horning, K., Seager, S., \& Harrington, J., 2007, Nature, 445, 892
\reference{Rowe} Rowe, J. F., et al., arXiV:0711.4111
\reference{Russ} Russell, H. N., 1916, ApJ, 43, 173
\reference{DS} Saumon, D., et al., 1996, ApJ, 460, 993
\reference{SRH} Seager, S., Richardson, L. J., Hansen, B. M. S., Menou, K., Cho, J. \& Deming, D., 
2005, ApJ, 632, 1122
\reference{SS00} Seager, S. \& Sasselov, D. D., 2000, ApJ, 502, L157
\reference{SWS} Seager, S., Whitney, B. \& Sasselov, D. D., 2000, ApJ, 540, 504
\reference{SG} Showman, A. P., \& Guillot, T., 2002, A\&A, 385, 166
\reference{SBH} Sudarsky, D., Burrows, A. \& Hubeny, I., 2003, ApJ, 588, 1121
\reference{VM03} Vidal-Madjar, A.,  Lecavelier des Etangs, A., D\'{e}sert, J.-M., Ballester, G. E., Ferlet, R., H\'{e}brard, G. \& Mayor, M, 2003, Nature, 422, 143
\reference{Will} Williams, P., Charbonneau, D., Cooper, C., Showman, A. \& Fortney, J., 2006, ApJ, 649, 1020
\end{references}
\end{document}